\documentclass[a4paper,fleqn,usenatbib]{mnras}

\usepackage{newtxtext,newtxmath}

\usepackage[T1]{fontenc}
\usepackage{ae,aecompl}

\usepackage{graphicx}  
\usepackage{amsmath}   
\usepackage{amssymb}   
\usepackage{subfig}
\usepackage{braket}

\newcommand{\gccm}{\,g cm$^{-3}$}

\newcommand{\kms}{\,km s$^{-1}$}

\newcommand{\half}{\frac{1}{2}}
\newcommand{\dext}{\rho_\text{ext}}
\newcommand{\pext}{p_\text{ext}}
\newcommand{\Mratio}{\frac{\left(1+\mathcal{M}_t^2\right)}{\left(1+\mathcal{M}_a^2\right)}}
\newcommand{\fcyl}{f_\text{cyl}}

\title[Accretion driven turbulence in filaments]{Accretion driven turbulence in filaments II: Effects of self-gravity}

\author[S. Heigl et al.]{
S. Heigl,$^{1,2}$\thanks{E-mail: heigl@usm.lmu.de}
M. Gritschneder,$^{1}$
A. Burkert,$^{1,2}$
\\
$^{1}$Universit\"ats-Sternwarte, Ludwig-Maximilians-Universit\"at M\"unchen, Scheinerstr. 1, 81679 Munich, Germany\\
$^{2}$Max-Planck Institute for Extraterrestrial Physics, Giessenbachstr. 1, 85748 Garching, Germany\\
}

\date{Accepted XXX. Received YYY; in original form ZZZ}

\pubyear{2018}

\begin{document}
\label{firstpage}
\pagerange{\pageref{firstpage}--\pageref{lastpage}}
\maketitle

\begin{abstract}

  We extend our previous work on simulations with the code \textsc{ramses}
  on accretion driven turbulence by including self-gravity and study the
  effects of core formation and collapse. We show that radial accretion
  onto filaments drives turbulent motions which are not isotropic but
  radially dominated. In contrast to filaments without gravity, the velocity
  dispersion of self-gravitating filaments does not settle in an equilibrium.
  Despite showing similar amounts of driven turbulence, they continually
  dissipate their velocity dispersion until the onset of core formation.
  This difference is connected to the evolution of the radius as it determines
  the dissipation rate. In the non-gravitational case filament growth is not
  limited and its radius grows linearly with time. In contrast, there is a
  maximum extent in the self-gravitational case resulting in an increased
  dissipation rate. Furthermore, accretion driven turbulence
  shows a radial profile which is anti-correlated with density. This
  leads to a constant turbulent pressure throughout the filament. As the
  additional turbulent pressure does not have a radial gradient it does not
  contribute to the stability of filaments and does not increase the
  critical line-mass. However, this radial turbulence does affect the radius
  of a filament, adding to the extent and setting its maximum value.
  Moreover, the radius evolution also affects the growth timescale of
  cores which compared to the timescale of collapse of an accreting
  filament limits core formation to high line-masses.

\end{abstract}




\section{Introduction}
\label{sec:introduction}

   Turbulent motions are a key feature in the highly complex dynamics of
   the interstellar medium (ISM) as demonstrated by the famous "Larson's
   Laws" \citep{larson1981}. Line observations of molecular clouds show a
   direct correlation between size and molecular linewidth which is usually
   interpreted as a consequence of a Kolmogorov-like turbulent cascade from
   supersonic motions on the scale of a tens of parsec sized molecular cloud
   down to the sonic point on the scale of parsec sized filaments
   \citep{kolmogorov1941, kritsuk2013, federrath2016, padoan2016}. As
   density structures are formed by the collision of supersonic flows in
   the turbulent cascade, the transition from supersonic to subsonic
   motions is essential for setting the scale at which turbulence stops
   to dominate and the first subsonic density structures form.

   Dust observations show that filamentary structure is ubiquitous on
   parsec sized scales in star-forming, as well as quiescent, molecular clouds
   and directly associated with core formation in filaments with
   supercritical line-masses \citep{andre2010,arzoumanian2011,andre2014}.
   The fact that these supercritical filaments show an increasingly
   supersonic internal velocity dispersion for larger line-masses, has
   been interpreted as a consequence of their gravitational collapse
   \citep{arzoumanian2013}. However, molecular line observations have shown
   that some filaments are actually made up of bundles of velocity coherent
   subcomponents in the line-of-sight velocity called fibres
   \citep{hacar2013, hacar2018} which also form in numerical simulations
   \citep{smith2014, moeckel2015, clarke2017}.
   While these subcomponents show trans- or subsonic linewidths, here
   relative motions create supersonic linewidths in spectrally low
   resolved superposition. Moreover, filaments which do not show
   any substructure are observed to be inherently subsonic
   \citep{hacar2011, hacar2016}.

   Two methods have been proposed to
   explain the formation of fibres. On the one hand, simulations by
   \citet{smith2016} showed that the subcomponents form first in collapsing
   clouds as a consequence of the turbulent cascade and are collected into
   individual filaments by large scale motions. This process is known as the
   "fray and gather" scenario. One the other hand, in the "fray and fragment"
   scenario proposed by \citet{tafalla2015}, the formation of fibres is
   explained by the sweep-up of residual motions inside a filament itself.
   Together with gravity, these motions concentrate material into subsonic
   velocity coherent entities in which core fragmentation takes place.
   \citet{clarke2017} pointed out that this model relies strongly on
   the vorticity of the gas in order to work. If the vorticity of two nearby
   regions is anti-parallel, the resulting flow compresses gas to the
   interface between the regions and thus leads to an overdense fibre.

   As indicated by \citet{hacar2016_2}, there are two
   distinct modes of turbulence. On the one hand, there is the classical
   microscopic turbulence where the density and velocity fields are
   continuous and isotropic as in the Kolmogorov model. On the other
   hand, the macroscopic turbulence observed in filaments containing fibres
   is created by multiple discreet overdensities with internal velocity
   dispersions of about the sound speed moving with supersonic motions
   relative to each other. However, care has to taken in interpreting the
   observations of fibres. As noted by \citet{zamora2017} and
   \cite{clarke2018}, velocity coherent structures in
   position-position-velocity space do not necessarily represent physical
   density structures in position-position-position space.

   Nevertheless, in the absence of a driving source, turbulent motions
   decay on the timescale of a crossing time
   \citep{maclow1998,stone1998,padoan1999,maclow1999,maclow2004}. This
   timescale can be very short for filaments if one assumes the driving
   scale is given by the filament diameter. In a first study
   (\citet{heigl2018_1}, hereafter called paper I), we used an external
   accretion flow motivated by the filaments self-gravity to provide a
   driving source of turbulence. While we explored the effects of accretion
   driven turbulence on non self-gravitating filaments, we now take
   self-gravity and core formation into account. Observationally,
   accreting material is expected to flow along striations, weak
   filamentary density enhancements perpendicular to the filaments and
   aligned with the magnetic field, and have accretion rates of the order
   of $10-100$ M$_\odot$ pc$^{-1}$ Myr$^{-1}$ \citep{palmeirim2013, cox2016}
   and infall velocities of the order of $0.25-1.0$ \kms
   \citep{kirk2013, palmeirim2013}. Independent of their formation process,
   we show that as long as a filament is embedded in surrounding material,
   self-gravity leads to a continuous inflow onto the filament which causes
   the creation of turbulent motions.

   In the following sections, we first introduce the basic concepts that we
   apply to our model (\autoref{sec:concepts}). Thereafter, we discuss
   the numerical set-up of our simulations (\autoref{sec:numericalsetup}).
   Then we present our findings (\autoref{sec:simulations}) and
   discuss the implications for core formation (\autoref{sec:core}).
   Finally, we summarize our findings (\autoref{sec:discussion}).

\section{Basic concepts}
\label{sec:concepts}

   This section presents the fundamental principles which we use to derive
   our models. We discuss the theoretical hydrostatic profile of filaments
   and how filaments behave in an ambient medium. Then we derive expected
   accretion rates onto filaments motivated by their self-gravity and how
   the accretion affects their radii. Finally, we present a simple model of
   how accreted kinetic energy is transformed into turbulent velocities
   and how they are able to add pressure support.

   \subsection{Hydrostatic equilibrium}

   We use the isothermal and hydrostatic equilibrium model of a filament
   with a density profile described by \citet{stodolkiewicz1963} and
   \citet{ostriker1964}:
   \begin{equation}
     \rho(r) = \frac{\rho_c}{\left(1+\left(r/H\right)^{2}\right)^{2}}
   \end{equation}
   where $r$ is the cylindrical radius and $\rho_c$ is its central density.
   The radial scale height $H$ is given by the term:
   \begin{equation}
      H^2 = \frac{2c_s^2}{\pi G \rho_c}
      \label{eq:height}
   \end{equation}
   where $c_s$ is the isothermal sound speed and $G$ the gravitational
   constant. For our simulations we assume that the isothermal gas has a
   temperature of $10 \text{ K}$. With a molecular weight of $\mu=2.36$ the
   isothermal sound speed is $c_s=0.19$ \kms. The critical line mass above
   which a filament will collapse under its self-gravity is calculated by
   integrating the profile to $r \rightarrow \infty$:
   \begin{equation}
      \left(\frac{M}{L}\right)_\text{crit} =
      \frac{2c_s^2}{G}\approx1.06\cdot 10^{16}
      \text{g cm}^{-1}\approx16.4 \text{ M}_{\sun}\text{ pc}^{-1}.
      \label{eq:lmcrit}
   \end{equation}
   Filaments with a lower line-mass than the critical value expand as long
   as there is no outside pressure. If one now assumes the filament is
   embedded in an ambient medium with a source of external pressure
   $p_\text{ext}$, the filament radius and line-mass are limited and one
   can introduce the parameter $\fcyl$ which is a measure of how close to
   the critical value the filament is:
   \begin{equation}
     \fcyl = \left(\frac{M}{L}\right)/\left(\frac{M}{L}\right)_\text{crit}.
   \end{equation}
   It varies from 0 for a non-existing filament to 1, where a filament has
   exactly its critical line-mass. As shown by \citet{fischera2012}, the
   radius $R$ of the filament is then given by the position where the internal
   pressure matches the external pressure and can be expressed as:
   \begin{equation}
     R = H \left(\frac{\fcyl}{1-\fcyl}\right)^{1/2}.
   \end{equation}
   Together with the relation between the central density $\rho_c$ and
   the density on the boundary of the filament,
   \begin{equation}
     \rho_c = \frac{\rho(R)}{(1-\fcyl)^{2}},
   \end{equation}
   one can write the radius as function of the boundary density:
   \begin{equation}
     R = \left(\frac{2c_s^2}{\pi G \rho(R)}
         \left(\fcyl(1-\fcyl)\right)\right)^{1/2}.
     \label{eq:rad}
   \end{equation}
   For a given boundary density, the radius has a maximum at $\fcyl=0.5$
   with a symmetric drop-off to zero at $\fcyl=0.0$ and $\fcyl=1.0$. The
   boundary density $\rho(R)$ depends on the outside pressure and thus on the
   mechanism responsible for the pressure. One possible source of pressure
   is the thermal pressure of the surrounding medium itself. If the density
   on the outside of the surface is given by $\dext$ and the medium can be
   assumed to be isothermal with sound speed $c_s$, the pressure acting
   onto the surface is:
   \begin{equation}
     \pext = \dext c_s^2.
     \label{eq:thermalp}
   \end{equation}
   If both, the filament and the surrounding gas, have the same temperature
   then the boundary density $\rho(R)$ and the density of the surrounding
   gas are the same. Different temperatures however, lead to a jump in
   density in order to establish hydrostatic equilibrium.

   \subsection{Filament accretion}

   Another possible source of outside pressure is accretion of material
   onto the filament. We are particularly interested in accretion due
   to the gravitational potential of the filament itself. The gravitational
   acceleration of a cylindrical distribution of mass with a given
   line-mass $M/L$ is \citep{heitsch2009}:
   \begin{equation}
      a = -\frac{2GM/L}{r}.
   \end{equation}
   The potential energy that a gas parcel with mass $m$ loses in
   free-fall when starting with zero velocity at a distance $R_0$ and
   accreting to the filament radius $R$ is given by integrating the
   acceleration over $r$:
   \begin{equation}
     E_\text{pot} = 2 G m (M/L) \ln\left(\frac{R_0}{R}\right)
   \end{equation}
   This leads to the accretion velocity $v_a$ at the surface $R$ in the
   case of cylindrical free-fall:
   \begin{equation}
   \begin{split}
     v_a &= 2\sqrt{G(M/L)\ln\left(\frac{R_0}{R}\right)} \\
     & = 1.1 \text{ km s}^{-1} \left(\frac{M/L}{16.4 \text{ M$_\odot$ pc$^{-1}$}}\right)^{1/2} \left(\frac{\ln\left(R_0/R\right)}{\ln(100)}\right)^{1/2}.
   \end{split}
   \end{equation}
   Assuming a reasonable filament with a radius of order 0.1 pc
   and a large region of gravitational influence on the scale of a
   molecular cloud (a factor of a hundred times its own radius) allows
   us to estimate an upper limit on the accretion velocity.
   A filament at 10 K needs a line mass several times larger than
   the critical line-mass to achieve an inflow velocity of even
   Mach 10.0. In our simulations, we set the inflow velocity to a fixed
   value at the inflow boundary. As the simulated domain is relatively
   small compared to the filament itself and the filament is not massive
   enough to accelerate accreting gas over the time it takes to reach
   the filament, we can assume a constant accretion rate set by the
   radius of the inflow region $R_0$ and the density at that radius
   $\rho_0$:
   \begin{equation}
     \frac{\dot{M}}{L} = 2\pi \rho_0 R_0 v_a.
     \label{eq:mdot}
   \end{equation}
   This leads to a time-independent density profile outside of the
   filament,
   \begin{equation}
     \rho(r) = \rho_0 \frac{R_0}{r}
     \label{eq:rhoext}
   \end{equation}
   with the outside density at the filament radius $R$ being
   \begin{equation}
     \dext = \rho_0 \frac{R_0}{R}.
   \end{equation}
   If the inflow onto the filament is fast enough, ram pressure will
   dominate over the isothermal boundary pressure. Comparing to
   \autoref{eq:thermalp}, this is the case if the inflow velocity is
   greater than the isothermal sound-speed. The total external pressure is
   then given by
   \begin{equation}
     \pext = \dext\left(c_s^2 + v_a^2\right) = \frac{\rho_0 R_0 c_s^2\left(1+\mathcal{M}_a^2\right)}{R}.
     \label{eq:pwg}
   \end{equation}
   where $\mathcal{M}_a$ is the Mach number of the accretion flow. The
   external pressure is balanced by the internal pressure of the filament at
   the position of the boundary. In addition to the thermal pressure,
   turbulent motions inside the filament could be able to exert an additional
   turbulent component, an assumption we test in this study. Together with
   a turbulent component, the pressure equilibrium can then be written as:
   \begin{equation}
     \rho(R)c_s^2\left(1 + \mathcal{M}_t^2\right) = \frac{\rho_0 R_0 c_s^2\left(1+\mathcal{M}_a^2\right)}{R},
   \end{equation}
   where $\mathcal{M}_t$ is the Mach number of the turbulent motions within
   the filament at the boundary. Solving for the boundary density and
   inserting the result into \autoref{eq:rad} gives the radius of a
   filament with an additional accretion pressure and internal turbulent
   motions as:
   \begin{equation}
     R = \frac{2c_s^2 \left(1+\mathcal{M}_t^2\right)}{\pi \rho_0 R_0 G \left(1+\mathcal{M}_a^2\right)} \left(\fcyl(1-\fcyl)\right).
     \label{eq:t_radwg}
   \end{equation}
   Although the radius evolution has now lost its dependence on the
   square root, the general shape of the curve remains unchanged. There
   still is a maximum at $\fcyl=0.5$ which only differs in its maximum
   value. Note that for $G \rightarrow 0$, \autoref{eq:t_radwg} transforms
   to the non-gravitational counterpart of \autoref{eq:radwogturb}.

   \subsection{Accretion driven turbulence}

   The analytical prediction of accretion driven turbulence is based on the
   energy budget of accreted kinetic energy being converted to turbulent
   energy and its subsequent dissipation. Following \citet{elmegreen2010}
   and \citet{klessen2010}, the change in turbulent energy $\dot{E}_t$ is
   given by the energy accretion rate $\dot{E}_a$ and the energy
   dissipation $\dot{E}_d$:
   \begin{equation}
     \dot{E}_t = \dot{E}_{a} - \dot{E}_d = (1-\epsilon)\dot{E}_{a}.
     \label{eq:eequ}
   \end{equation}
   The energy accretion rate is given by the accreted kinetic energy
   \begin{equation}
     \dot{E}_{a} = \half \dot{M} v_a^2
   \end{equation}
   and the energy loss through dissipation by
   \begin{equation}
     \dot{E}_d \approx \frac{E_{t}}{\tau_d}= \half \frac{M\sigma^3}{L_d},
     \label{eq:ediss}
   \end{equation}
   where the turbulent energy is expected to decay on the timescale of a
   crossing time:
   \begin{equation}
     \tau_d \approx \frac{L_d}{\sigma}.
   \end{equation}
   \citet{elmegreen2010} also introduce the efficiency factor $\epsilon$ as
   fraction of accreted energy used to sustain the turbulent motions:
   \begin{equation}
     \epsilon = \left|\frac{\dot{E}_d}{\dot{E}_a}\right|.
     \label{eq:epsilon}
   \end{equation}
   \citet{heitsch2013} used this approach together with a driving scale
   of the filament diameter $L_d = 2R$ to calculate the velocity dispersion
   in dependence of inflow velocity:
   \begin{equation}
     \sigma = \left(2\epsilon R(t) v_a^2\frac{\dot{M}}{M(t)}\right)^{1/3}.
     \label{eq:heitsch}
   \end{equation}
   For a linear evolution in time of the radius and mass, this relation
   predicts a constant level of velocity dispersion which is determined
   by the inflow velocity. While our simulations do find an equilibrium
   in velocity dispersion, we cannot match the scaling of the prediction.
   In contrast, our simulations show a linear relation of the density
   weighted velocity dispersion and inflow velocity as shown in paper I.
   Note, that any model for the scaling of velocity dispersion and
   inflow velocity or even an equilibrium is directly tied to the
   evolution of the radius. This is due to the fact that the energy
   dissipation rate depends on the crossing time. Rewriting the
   energy dissipation rate as
   \begin{equation}
     \dot{E}_d = \half \frac{M\sigma^3}{L_d}
               = \half \frac{\dot{M}\sigma^3 t}{2 R(t)},
   \end{equation}
   all terms in \autoref{eq:eequ} depend on the mass accretion rate and
   can be simplified to
   \begin{equation}
     \sigma^2 = \alpha v_a^2 - \frac{\sigma^3 t}{2 R(t)}.
     \label{eq:equilibrium}
   \end{equation}
   Here we introduce the factor $\alpha$ to account for energy losses
   in the isothermal oblique accretion shocks at the surface of the
   filament where the turbulent motions are created. In order to reach
   an equilibrium velocity dispersion, all terms must be independent of
   time. This is only the case under two conditions. Either the radius
   grows linear in time as then the dissipation rate is constant
   or the radius evolves superlinear in time for which the last term
   vanishes at large timescales. This explains why we have an
   equilibrium in the non self-gravitational case as we find a
   linear time evolution of the radius. We revisit the equilibrium level
   and give an explanation for the scaling in the appendix. In the case
   including self-gravity discussed in this
   paper, the radius evolves as a complicated function of time
   (\autoref{eq:rad}) and core formation could impact the velocity
   dispersion substantially.

   \subsection{Effect of turbulence on radial stability}

   An important question that we want to answer is if turbulence can
   increase the stability of filaments. Previous studies modeled the
   effects of turbulence on the equation of state either by
   logatropic models based on the scaling relations of molecular
   clouds \citep{larson1981} where $p_\text{turb} \sim \ln\rho$
   \citep{lizano1989, 1gehman1996, 2gehman1996, mclaughlin1997, 1fiege2000},
   or negative index polytropes, where the polytropic exponent is between
   zero and one, as is the case for Alf\'{e}nic turbulence
   \citep{maloney1988, fatuzzo1993, mckee1995}. One could also assume
   that isotropic turbulence behaves as an additional component to the
   thermal pressure and add it to the scale height:
   \begin{equation}
     H^2 = \frac{2c_s^2\left(1+\mathcal{M}_t^2\right)}{\pi G \rho_c}.
   \end{equation}
   Integrating over the filament profile as done for \autoref{eq:lmcrit},
   leads to an adjustment of the critical line-mass:
   \begin{equation}
     \left(\frac{M}{L}\right)_\text{crit} =
     \frac{2c_s^2\left(1 + \mathcal{M}_t^2\right)}{G}.
     \label{eq:lmturb}
   \end{equation}
   Thus, turbulence would be able to increase the maximum line-mass
   a filament could sustain. Note that this adjustment does not
   affect the formula for the radius in \autoref{eq:t_radwg} as the
   additional turbulent term not only enters directly over the scale
   height, but also over $\fcyl$ and thus cancels out.
   However, an increased maximum line-mass would lead to an offset
   to the point of the maximum radial extent from the value 0.5 if
   plotted against the unadjusted maximum thermal line-mass. In
   addition, an increased stability would lead to a delayed collapse
   of cores in a filament as long as they do not dissipate their
   turbulence on a much faster timescale. Therefore, we will also
   directly test the impact of turbulence by investigating the
   radial evolution and core collapse in filaments.

\section{Numerical set-up}
\label{sec:numericalsetup}

   All our simulations were executed with the code \textsc{ramses}
   \citep{teyssier2002} which uses a second-order Godunov scheme
   to solve the conservative form of the discretised Euler equations
   on an Cartesian grid. For our runs we applied the MUSCL
   scheme (Monotonic Upstream-Centred Scheme for Conservation Laws,
   \citet{vanLeer1977}) together with the HLLC-Solver
   (Harten-Lax-van Leer-Contact \citep{toro1994}) and the
   multidimensional MC slope limiter (monotonized central-difference
   \citep{vanLeer1979}).

   \begin{figure*}
     \includegraphics[width=2.0\columnwidth]{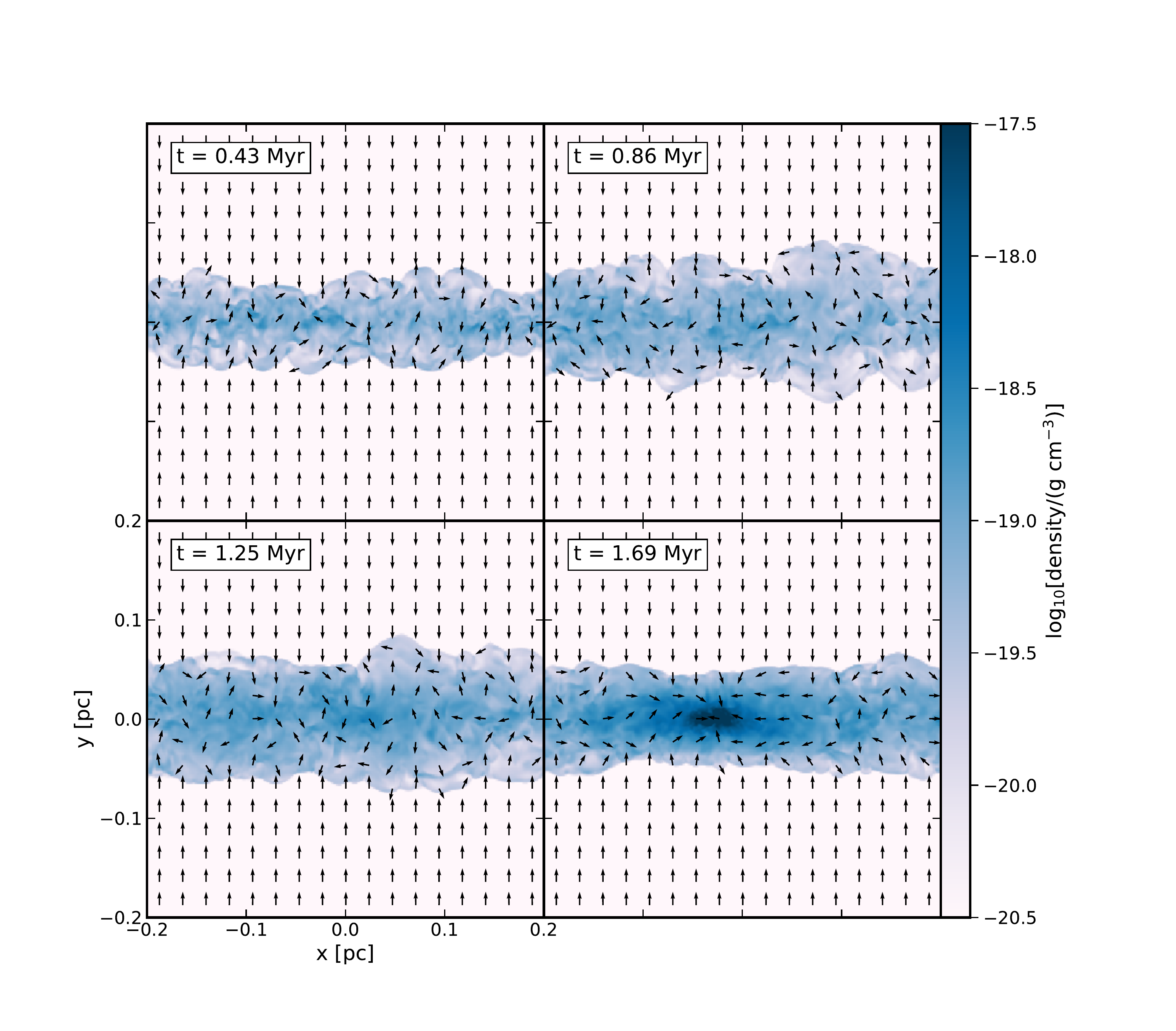}
     \caption{Evolution of an accreting filament with an inflow velocity
     of Mach 6.0 and an accretion rate of
     $8.4\text{ M$_\odot$ pc$^{-1}$ Myr$^{-1}$}$. The image shows four
     slices at different evolutionary times through the centre of the
     filament together with the directions of the local motion. In
     order to improve the contrast, we excluded the ambient medium from the
     colour bar. The cylindrical accretion induces obvious turbulent
     motions until a core is formed where the dominant velocity pattern
     changes to an accretion onto the core.}
     \label{fig:slice}
   \end{figure*}

   Our simulations cannot resolve the evolution of the molecular cloud
   and the detailed velocity dispersion inside the filament at the
   same time. Therefore we focus on a preset converging radial
   flow onto a self-gravitating, isothermal filament in the centre of the
   box. We use a 3D box with a periodic boundary condition in the
   x-direction and outflow boundaries in the other two directions. As
   \textsc{ramses} has no radial boundary we define a cylindrical inflow
   zone with a radius of the boxsize and a thickness of two cells from which
   we drive a radial inflow onto the central x-axis of the box. The
   inflow zone has a fixed density and inflow velocity which is
   continuously renewed every timestep.
   The inflow leads to a build-up of a filament with a radius which is
   limited by gravity. The periodic boundary prevents the filament from
   collapsing along its axis and prohibits the loss of turbulent motions.
   As the radius does not grow to large values, we can often
   optimize the resolution of our simulations and use a boxsize of 0.4 pc
   which is half as large as the standard boxsize used in paper I.
   Therefore, in order to keep the initial conditions equivalent
   to paper I, we double the outer boundary density for the majority of our
   simulations to a value of $\rho_0 = 7.84\cdot10^{-22}$\gccm which
   corresponds to about $2 \cdot 10^2$ particles per cubic centimeters
   for a molecular weight of $\mu=2.36$. While we perform simulations for
   inflow Mach numbers ranging from 2.0 to 15.0, our analysis mainly
   concentrates on a reference case of a filament with an inflow velocity
   of Mach 6.0 or equivalently an mass accretion rate of
   $\dot{M}/L = 16.8 \text{ M$_\odot$ pc$^{-1}$ Myr$^{-1}$}$ as the general
   results do not change with inflow velocity. We vary the mass accretion
   rate onto the filament by adjusting the boundary density and for
   simulations where we show the equilibrium level of turbulence, we
   reduce the density in the inflow region to a value which gives enough
   time to allow the equilibrium to settle without the filament collapsing.
   We also set the initial density inside the domain to the outer boundary
   density and vary the density in each cell with a random perturbation of
   50\%. The gas is set to be isothermal with a temperature of 10 K and the
   cells surrounding the inflow zone are given the same constant density and
   do not affect the simulation.

   The minimum resolution is set to $256^3$, which at this boxsize is
   equivalent to the minimum resolution of about 0.002 pc used
   in paper I and guarantees that we resolve the filament across its diameter
   with a minimum of around 50 cells at all times. We employ adaptive
   mesh refinement (AMR) in order to resolve higher densities. Over the
   evolution of the simulation the filament stays unrefined and AMR only
   plays  a role in the high density cores. The maximum resolution is set to
   $512^3$ and in such a way that we terminate the
   simulation as soon as we do not fulfill the Truelove criterion for
   the maximum density within a factor of 16 in all simulations
   \citep{truelove1997}.

\section{Simulations}
\label{sec:simulations}

   In this section we analyse the simulations in detail. While we
   concentrate on an inflow Mach number of 6.0, we performed the same
   analysis for all simulations ranging from Mach 2.0 to 15.0 and the
   results are generally valid independent of inflow Mach number.
   As in paper I we study the velocity dispersion and radius of the filament.
   We show the typical evolution of a simulation in \autoref{fig:slice}
   where we plot slices through the central axis of the filament.
   As the filament grows in mass it first expands, reaches a maximum in
   radius and decreases again in radius. Towards reaching the critical
   line-mass of $\fcyl=1.0$, usually one or more cores can be seen to
   condense inside the filament. Our simulation ends when we reach the
   maximum allowed density due to the Truelove criterion of around
   $10^{-17}$\gccm or $10^7$ particles per cubic centimeter where we would
   need to insert a sink particle. As the subsequent core collapse happens
   on much shorter timescales than the evolution of the filament and we do
   not model expected feedback from protostellar outflows, we terminate our
   simulations as soon as we reach this threshold.

   In comparison with the study of a similar set-up by
   \citet{clarke2017} which used the Smooth Particle Hydrodynamic code
   \textsc{gandalf} \citep{hubber2018}, the density distribution inside
   the filament in our study is much smoother despite showing strong
   turbulent velocities. This is likely caused by the difference in initial
   conditions used by \citet{clarke2017} which includes an initial turbulent
   velocity field in the accreting flow leading to larger substructures in the
   accreted material. In contrast, the turbulence in our study is generated
   by oblique shocks on the surface of the filament due to small-scale
   inhomogeneities on a cell-to-cell basis of the random perturbation in the
   accretion flow. The outcome reflects the distinction between the
   macroscopic and microscopic turbulent model mentioned in the
   introduction as we do not observe the formation of fibre-like
   structures in our simulations but rather small-scale and continuous
   turbulent density enhancements.

   \subsection{Evolution of the velocity dispersion}

   \begin{figure}
     \includegraphics[width=1.0\columnwidth]{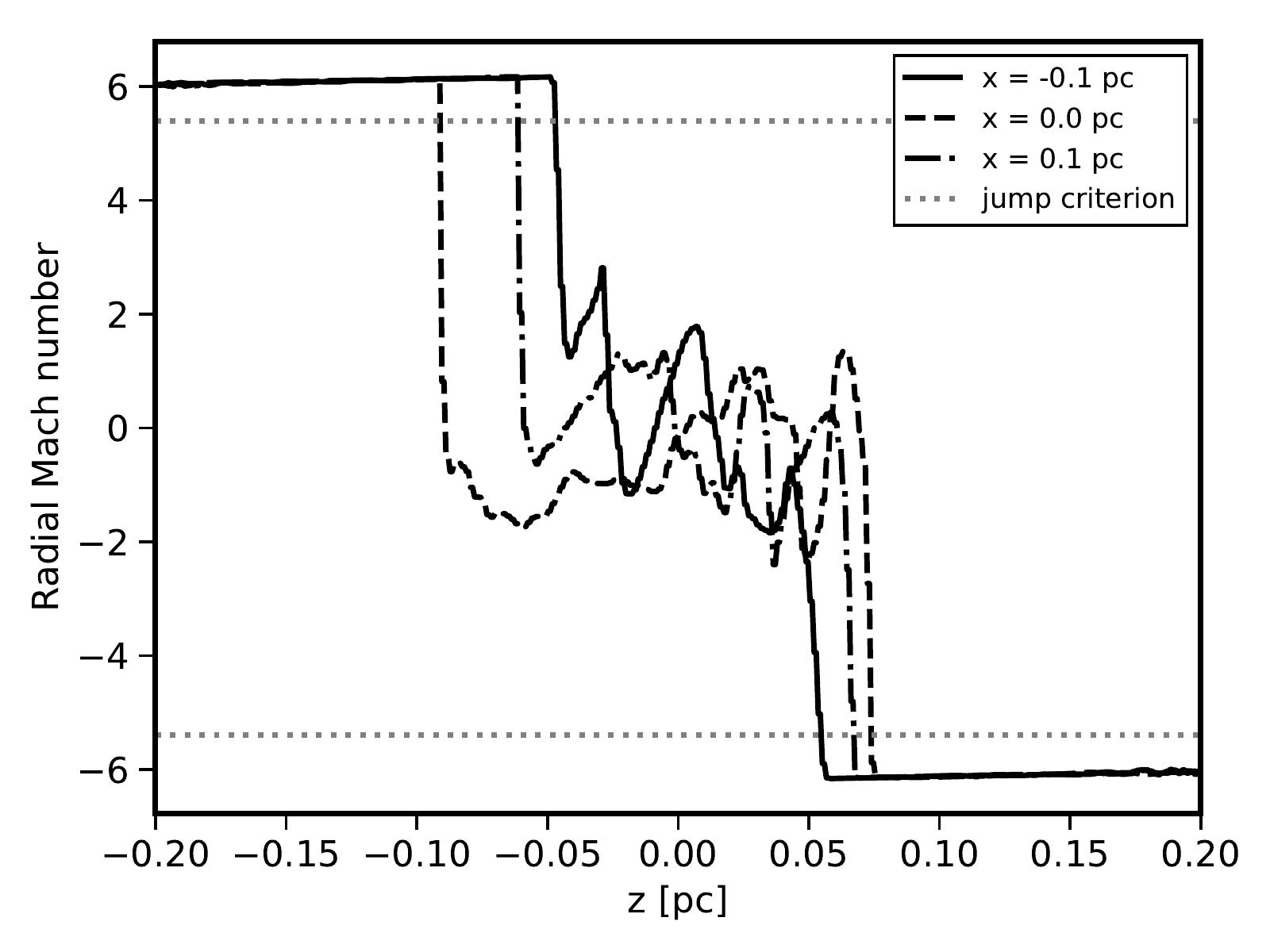}
      \caption{Example velocity cuts through the centre of the
      filament at different positions along the filament for a Mach 6.0
      inflow. We define cells belonging to the filament
      when the radial velocity with respect to the box-centre falls
      below 0.9 times the boundary inflow velocity. This value is shown
      by the dotted gray lines. As one can see there is a strong drop in
      radial velocity as soon as the accretion flow reaches the filament
      edge.}
      \label{fig:rad_vel}
   \end{figure}

   In order to calculate the velocity dispersion of the filament gas,
   we need to distinguish it from the accretion flow. Due to the formation
   of an accretion shock at the filament boundary, there is a clear jump in
   density and a drop in radial velocity of the material. However, the
   density inside the filament can vary considerably and therefore,
   instead of using an imprecise density jump criterion, we use the change
   in radial velocity as an indicator of filament material. We consider a
   cell to be part of the filament if its radial velocity with respect to
   the box centre has dropped to at least 0.9 of its boundary value as
   illustrated in \autoref{fig:rad_vel}. As the velocity jump is a very
   clear break, the calculated results do not depend strongly on the
   exact value of the velocity drop as it at most includes or excludes
   single cells which we confirm via a visual inspection of the filament
   material.

   For the determination of the velocity dispersion in paper I, we used the
   standard deviation of the density weighted velocity:
   \begin{equation}
     u_i = \frac{m_i v_i}{\left<m\right>}
         = \frac{N m_i v_i}{\sum m_i}
         = \frac{N m_i v_i}{M}
   \end{equation}
   where $N$ is the total number of cells and $\left<\right>$ indicates the
   mean of a distribution. A more common way of defining the velocity
   dispersion of gas with zero mean velocity is by calculating the total
   kinetic energy in the gas:
   \begin{equation}
     \sigma = \sqrt{\frac{1}{M}\sum_i m_i v_i^2}.
     \label{eq:ekin}
   \end{equation}
   In the limit that the density is constant, both ways of calculating the
   velocity dispersion are equivalent as the velocity mean is zero and the
   standard deviation then is given by
   \begin{equation}
     \sigma_u = \sqrt{\left<u^2\right>-\left<u\right>^2}
     = \sqrt{\frac{N^2\sum_i m_i^2 v_i^2}{N M^2}}
     = \sqrt{\frac{m_i\sum_i v_i^2}{M}}
   \end{equation}
   where we use the definition of the total mass $M=Nm_i$. As the density
   inside a filament is not constant in our simulations, we expect different
   measured values of the velocity dispersions for the different methods.
   Especially for large values of turbulence the differences should increase
   as more and more shocks form inside the filament with larger density
   contrasts. For this study we use the method of the total kinetic energy
   (\autoref{eq:ekin}).

   \begin{figure}
     \includegraphics[width=1.0\columnwidth]{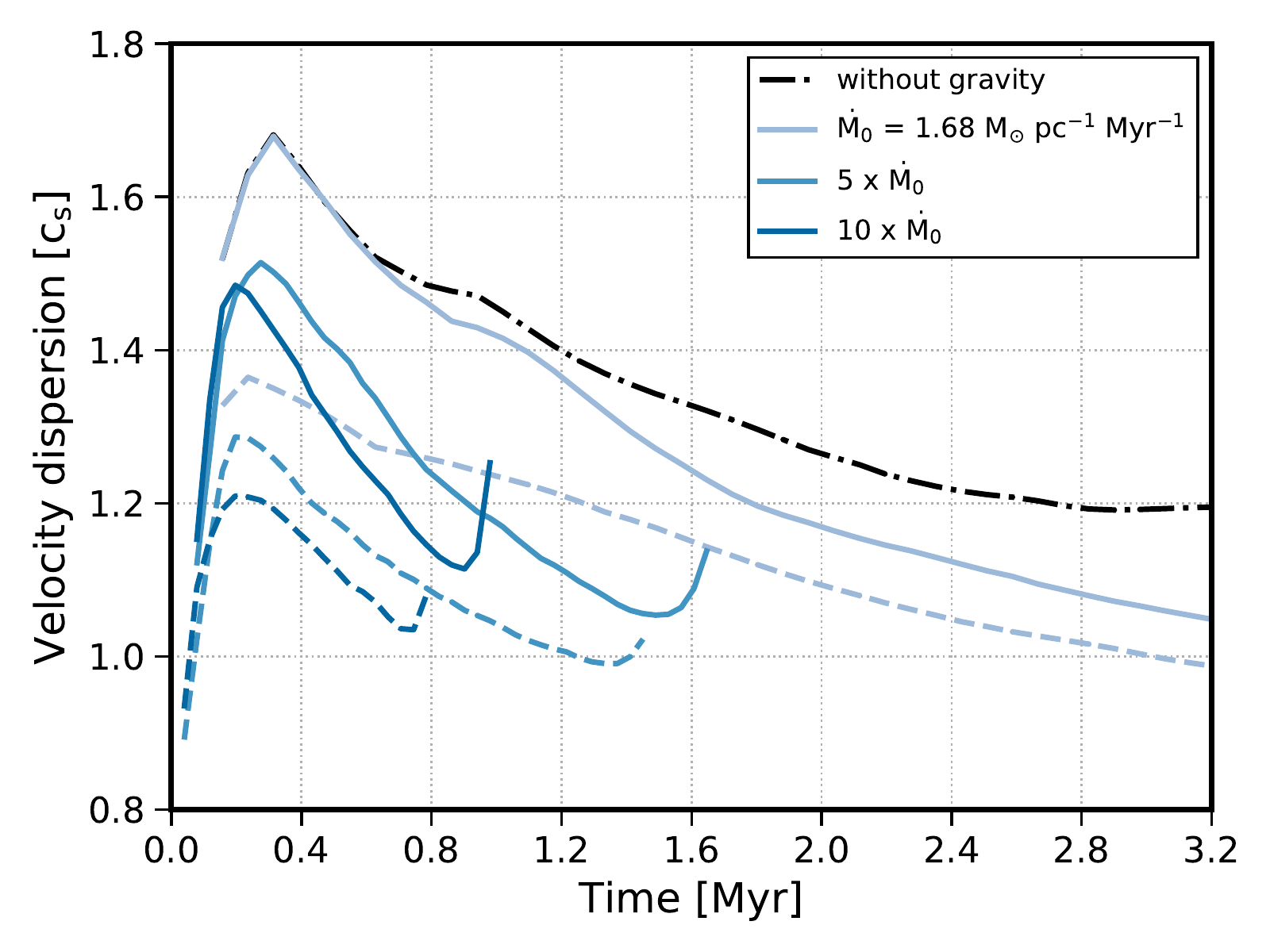}
      \caption{Evolution of the velocity dispersion for a Mach 6.0 inflow with
      varying accretion rate. We include a filament without gravity given
      by the dashed-dotted line as a reference case. The solid lines show
      filaments with different accretion rates. The dashed lines are the
      same initial values but with a pre-existing 1/r profile inside the
      box which smooths out the initial accretion shock.}
      \label{fig:t_sig}
   \end{figure}

   In paper I we observed that without self-gravity of the filament
   gas, the velocity dispersion settles to a constant equilibrium value over
   time. This behavior was also found in the study by \citet{clarke2017}
   which already included gravity. We want to test if it is possible to
   reproduce the transition to equilibrium if we include self-gravity in
   \textsc{ramses}. We measure the velocity dispersion of the filament gas
   for the same inflow velocity of Mach 6.0 but for different mass accretion
   rates and plot the evolution in \autoref{fig:t_sig}. The black
   dashed-dotted line is the evolution of the velocity dispersion of
   the non-gravitational case which has the same mass accretion rate
   as our reference case of 16.8 M$_\odot$ pc$^{-1}$. It ends in the same
   equilibrium value of an inflow velocity of Mach 6.0 presented in paper
   I.

   The solid light blue curve shows a filament with an accretion rate
   which is ten times lower than that of the reference case and the
   non-gravitational case. In contrast to the non-gravitational case, the
   velocity dispersion continuously drops off even if we continue the
   simulation to larger times. If we increase the mass accretion rate by
   a factor of five, as given by the medium blue line, or a factor of ten,
   as given by the dark blue line, the velocity dispersion drops off even
   faster and ends when core formation sets in where the velocity dispersion
   increases due to collapse motions onto the core. We use the high inflow
   rate case as our fiducial case for further analysis.

   We also test if our initial condition influences the evolution of the
   velocity dispersion. Therefore, we change our initial density profile in
   the box from a flat distribution to a $1/r$ profile, consistent with
   \autoref{eq:rhoext} as if the accretion flow has already been established.
   The corresponding lines are shown in \autoref{fig:t_sig} by the dashed
   lines. We see that it removes the initial spike in velocity
   dispersion. For higher accretion rates the initial value from where
   the velocity dispersion decays is lower but we also do not reach
   an equilibrium value.

   \begin{figure}
     \includegraphics[width=1.0\columnwidth]{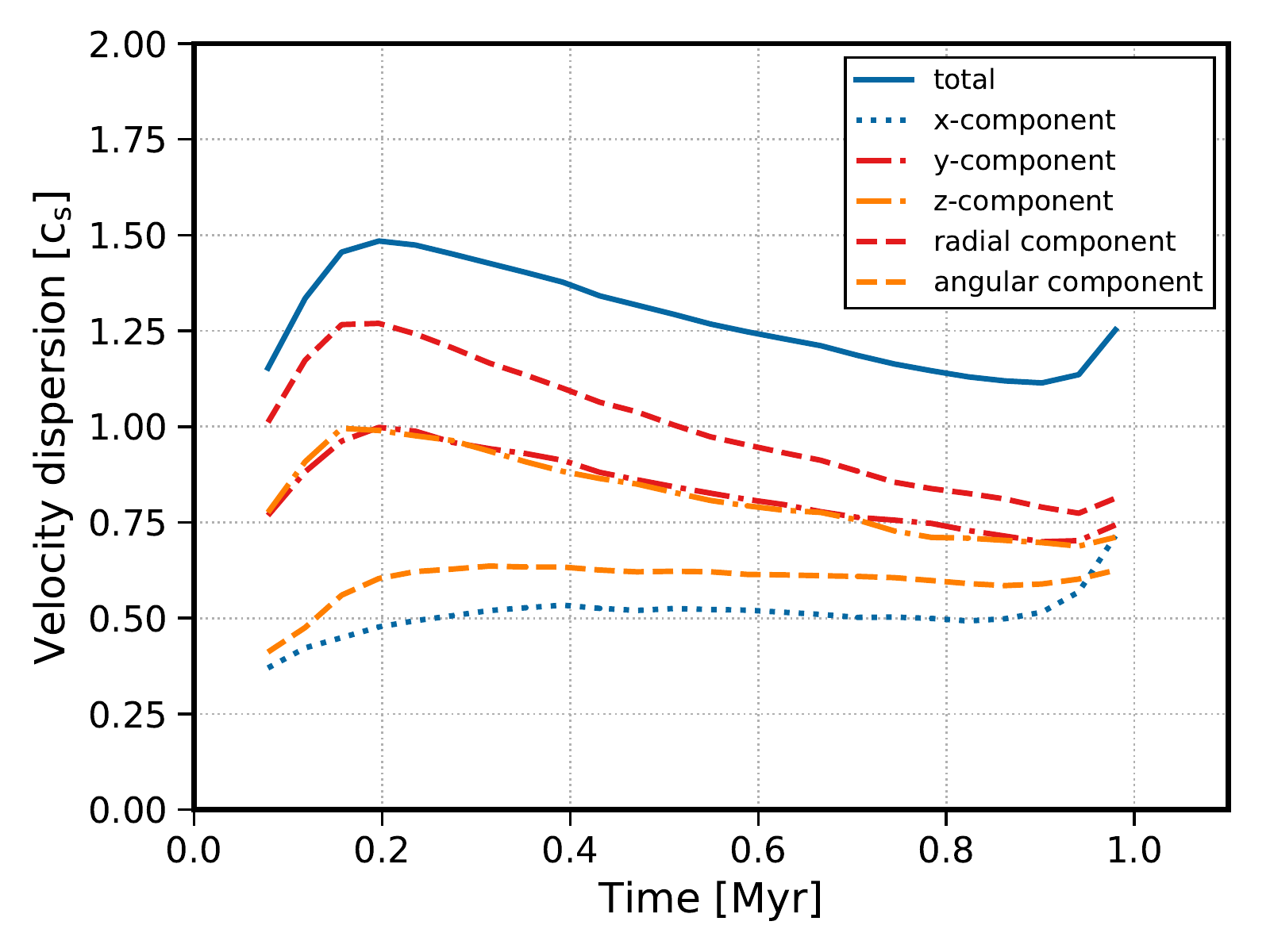}
     \caption{Evolution of the velocity dispersion of all components for a
     Mach 6.0 inflow with mass accretion rate of 16.8 M$_\odot$ pc$^{-1}$
     Myr$^{-1}$. The total velocity dispersion is shown by the blue solid
     line, the Cartesian components in y and z direction by the red and orange
     dashed-dotted lines and the cylindrical components in radial direction
     and in angular direction by the dashed red and orange lines respectively.
     The x-component, which is valid for both geometries is given by the
     blue dotted line.}
     \label{fig:t_ekin_split}
   \end{figure}

   We also split up the velocity dispersions into its cylindrical components
   by splitting the respective velocities into the x-component along the
   filament axis, the azimuthal and the radial velocities. Their evolution
   is shown in \autoref{fig:t_ekin_split} together with the Cartesian y and
   z component of the velocity dispersion. One can see that the turbulent
   motions are dominated by the radial velocity component and also only
   decay in the radial component while the other two stay constant over time.
   This is due to the difference in crossing times. Not only is the initial
   radial dispersion more than twice as large as the other components but
   it has also the lowest driving scale, with the azimuthal driving scale
   being a factor $\pi$ larger and the driving scale along the filament
   being the boxsize in theory.

   Therefore, albeit seeing an equilibrium velocity dispersion in
   filaments without gravity, gravitational collapse does influence the
   velocity dispersion by reducing it over time until the point of core
   collapse where collapse motions increase it again. This result is in
   agreement with \autoref{eq:equilibrium}, as for the non
   self-gravitational case, the radius evolves linearly, setting a constant
   dissipation rate and allowing for the velocity dispersion to settle to an
   equilibrium value. However, for self-gravitational filaments the radius
   has a maximum value and decreases again for large line-masses. This leads
   to a constant change in dissipation rate and thus no equilibrium can be
   achieved.

   \subsection{Radial evolution}

   \begin{figure}
     \includegraphics[width=1.0\columnwidth]{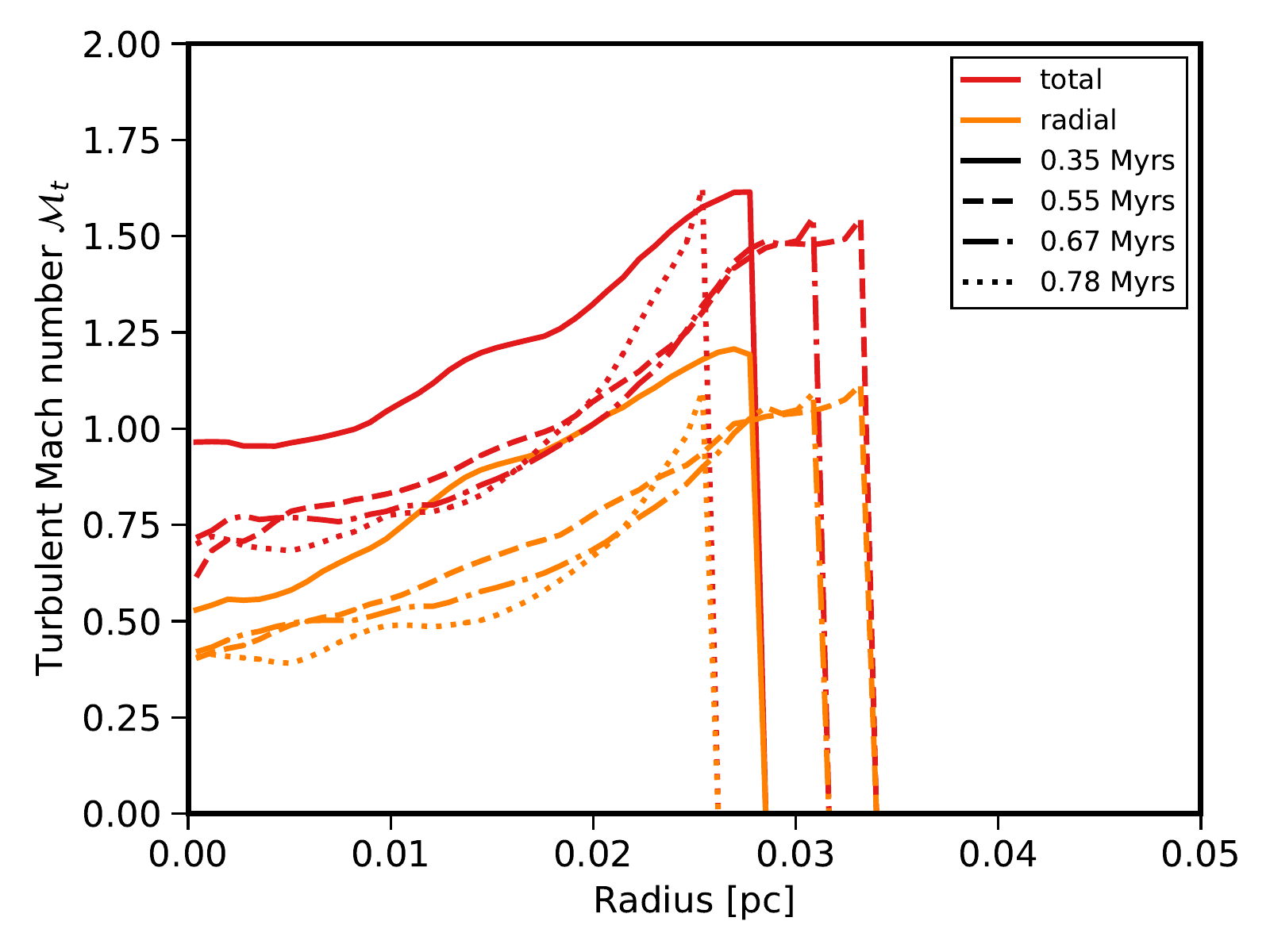}
     \caption{Turbulent Mach number averaged in radial bins of our fiducial
     filament with a Mach 6.0 accretion flow and an accretion rate of 16.8
     M$_\odot$ pc$^{-1}$ Myr$^{-1}$ measured at different timesteps
     given by the different linestyles. The total velocity dispersion is
     shown in and the radial velocity dispersion in orange.}
     \label{fig:rad_vkin}
   \end{figure}

   We showed that in the non self-gravitational case, the radius evolves
   linearly in time and is supported by the radial turbulent motions
   (see also Appendix A). According to \autoref{eq:rad}, the radial
   expansion is limited in the self-gravitational prediction, reaching a
   maximum at $\fcyl = 0.5$, followed by a subsequent decrease. In order to
   assess the impact of turbulent motions on the radial extent, we compare
   \autoref{eq:t_radwg} to the measured radius. If turbulence has an impact on
   the scale height, one would see an off-set in the radius maximum. This
   off-set can be substantial, e.g. a factor of two in the case where the
   velocity dispersion is larger than the order of the sound speed. Thus, by
   measuring the radial evolution, we can not only test the impact of
   turbulence on the radius but also if it acts as additional pressure
   support.

\begin{figure}
     \includegraphics[width=1.0\columnwidth]{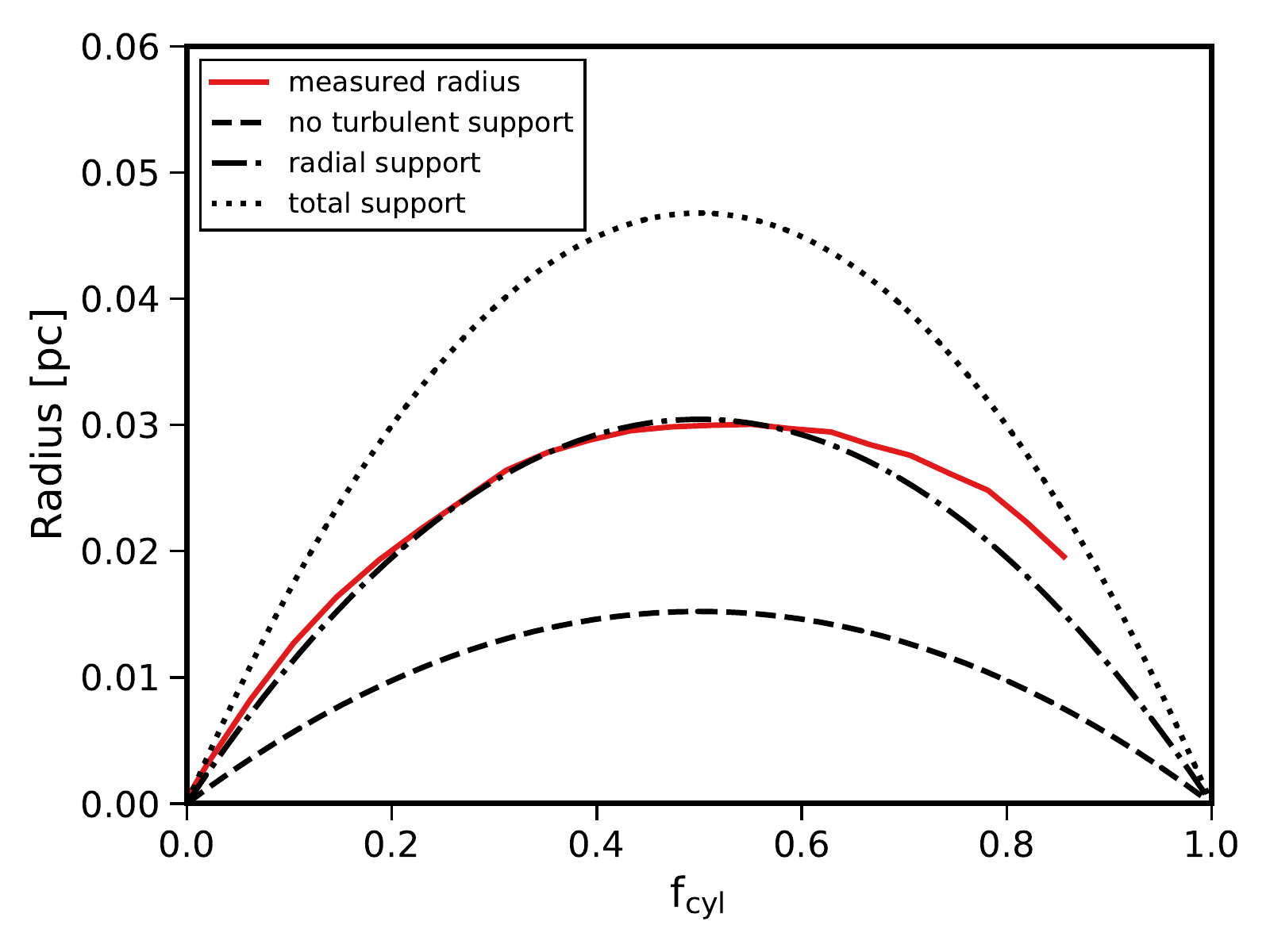}
     \caption{Evolution of the measured average radius of a filament with
     a Mach 6.0 accretion flow and an accretion rate of 16.8
     M$_\odot$ pc$^{-1}$ Myr$^{-1}$ compared to the average line-mass.
     The curve ends as soon as the core reaches the critical line-mass
     and collapses. The analytical evolution of the
     radius (\autoref{eq:t_radwg}) without turbulent pressure contribution
     is shown by the dashed line, with radial turbulence support as the black
     dashed-dotted line and with total turbulence support as the
     black dashed-dotted line.}
     \label{fig:f_rad}
   \end{figure}

   Before we compare the measured radius to \autoref{eq:t_radwg}, we have to
   determine the turbulent Mach number at the filament boundary, as
   $\mathcal{M}_t$ in the above equation is not the total Mach number, but
   the one determining hydrostatic equilibrium at the boundary. In
   order to do so, we have to calculate the radial profile of the turbulent
   Mach number. For large accretion rates the filament is too thin to
   determine a reasonable radial profile. Thus, we increase the resolution
   by re-simulating our fiducial case with a four times smaller box with a
   size of 0.1 pc.
   As the inflow region has a 1/r density profile, we also have to
   increase the inflow density by a factor of four in order to only simulate
   a zoomed-in sub-volume. This guarantees us the same inflow conditions and
   accretion rate on a smaller scale. We take each slice of
   the filament, split the domain in radial bins with a width of 4 cells
   and thus a physical bin size of $1.5\cdot10^{-3}$ pc,
   subtract the mean velocity of the bin, determine the total kinetic energy
   in the bin and finally average over all slices along the filament.
   The result is shown in \autoref{fig:rad_vkin}, where we plot both,
   the Mach number of the total velocity dispersion in
   red and the Mach number of the radial velocity in orange, for several
   different timesteps distinguished by the different linestyles. As one
   can see, the velocity dispersion is not constant throughout the filament,
   but is minimal at the filament centre and has its maximum at the boundary.
   This shows how turbulent motions are stirred at the surface of the
   filament and dissipate in the higher density layers. Over time,
   it is the high density interior which looses significantly in turbulent
   velocities. In order to calculate the predicted radius we need to use
   the boundary value. In principle, one would need to determine this
   value at every time step but we do not see a significant change of
   the boundary value and it stays close to constant over
   time. We use both boundary values, the total and the radial Mach number, in
   \autoref{eq:t_radwg} and plot the predicted radius as well as the
   average measured radius against the average line-mass in
   \autoref{fig:f_rad}. As its value varies along the filament as can
   be seen from \autoref{fig:rad_vel}, the radius itself is measured similar
   as the average of every slice along the filament.

   The measured radial
   evolution is shown by the solid red line, the case of no turbulent
   support as the dashed curve, the case of radial turbulent pressure
   support as the dashed-dotted line
   and the case of total turbulent pressure support as the dotted curve.
   The measured radius follows closely the radial turbulent support model.
   Note, that the same is true in the non self-gravity case in
   \autoref{fig:t_rad}. Only the radial motions contribute to the
   hydrostatic equilibrium. One can see that the curve reaches its
   maximum value at about $\fcyl=0.5$. It is important to note, that
   while turbulence does influence the maximum radius, it does not, or at
   most only marginally, affect the maximum line-mass and thus the point of
   where the radius reaches its maximum. According to \autoref{eq:lmturb},
   we would expect a maximum line-mass of twice the isothermal maximum
   mass given the velocity dispersion created by an inflow
   velocity of Mach 6.0. Thus, the radial evolution
   should peak at a unadjusted value of about $\fcyl=1.0$. Therefore,
   there is no indication of radial pressure support against
   gravity by turbulence as is consistent with the radial collapse
   discussed in the next section.

   \begin{figure}
     \includegraphics[width=1.0\columnwidth]{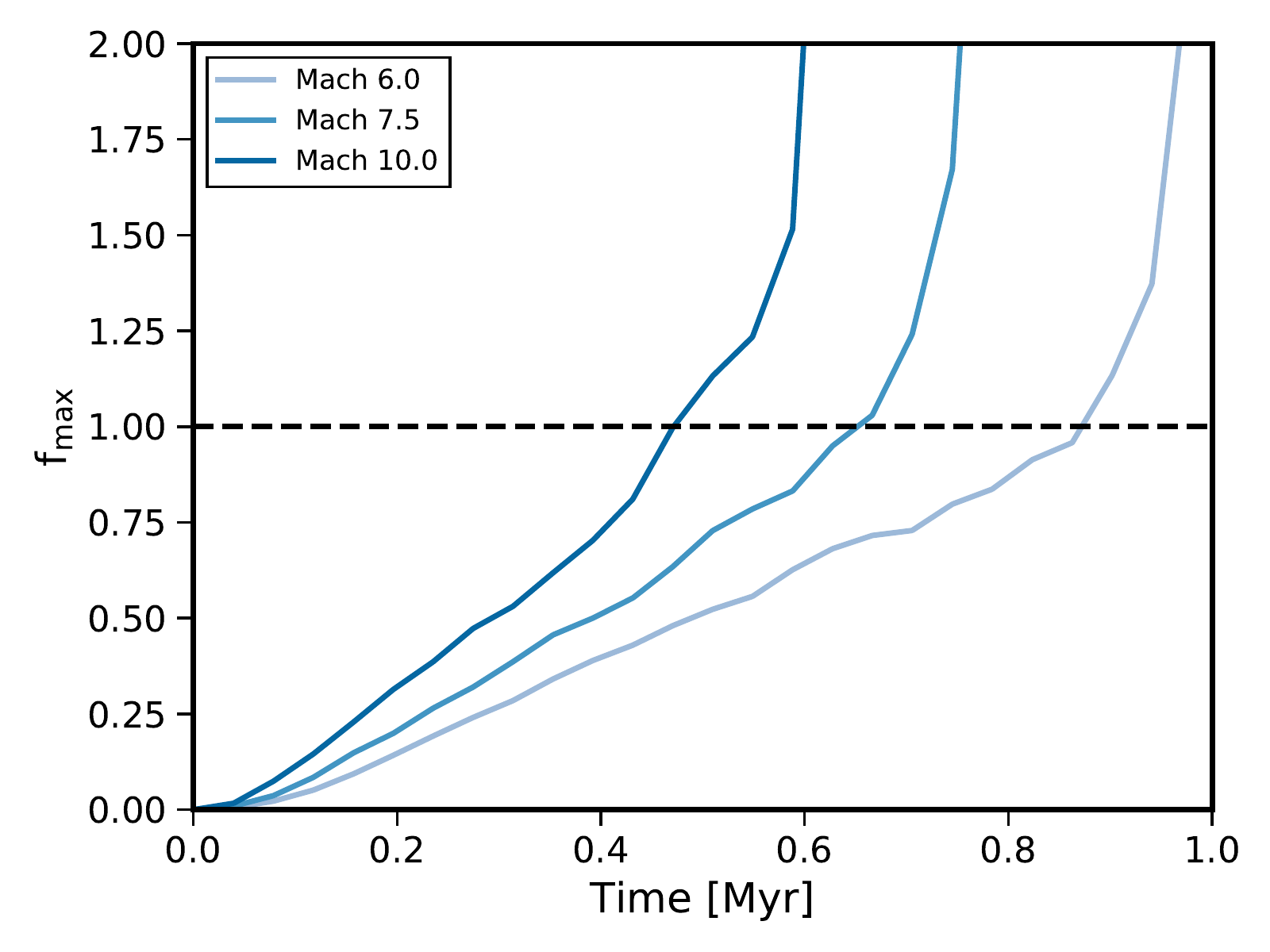}
     \caption{Time evolution of the maximum line-mass for different inflow
     Mach numbers. Due to the increasing mass in the filament, the maximum
     line-mass grows linearly. As soon as it reaches the unmodified
     critical line-mass at a value of 1.0, the core at this position
     collapses radially as can be seen in the non-linear evolution.}
     \label{fig:t_fmax}
   \end{figure}

   \subsection{Core collapse}

   We showed that turbulence does not have an effect on where the radial
   evolution has its maximum with respect to the filament line-mass.
   Therefore, we assume that there is no additional support against radial
   collapse which we test by analysing the growth of the forming core and
   whether or not it collapses at the critical line-mass. We plot the
   line-mass at the position of the core against time in
   \autoref{fig:t_fmax}. In \citet{heigl2016} we showed that the
   radial collapse of the core is visible in the non-linear evolution of
   the line-mass. If turbulence indeed plays a role for the stability of
   a filament, we expect an offset from $\fcyl = 1.0$ with respect to the
   line-mass growth change from linear to non-linear. If we adapt
   \autoref{eq:lmturb}, we predict a shift in the critical
   line-mass to at least double the usual value for turbulent Mach numbers of
   the same order as the sound speed. As one can see from the form of the
   curves, a non-linear evolution sets in as soon as the local line-mass
   at the position of the core exceeds the critical line-mass determined
   without turbulent support. This shows again that turbulence in our
   simulations does not have a supporting effect on the line-mass,
   consistent with the findings of the radial evolution.

   \subsection{Why is there no pressure support?}

\begin{figure}
     \includegraphics[width=1.0\columnwidth]{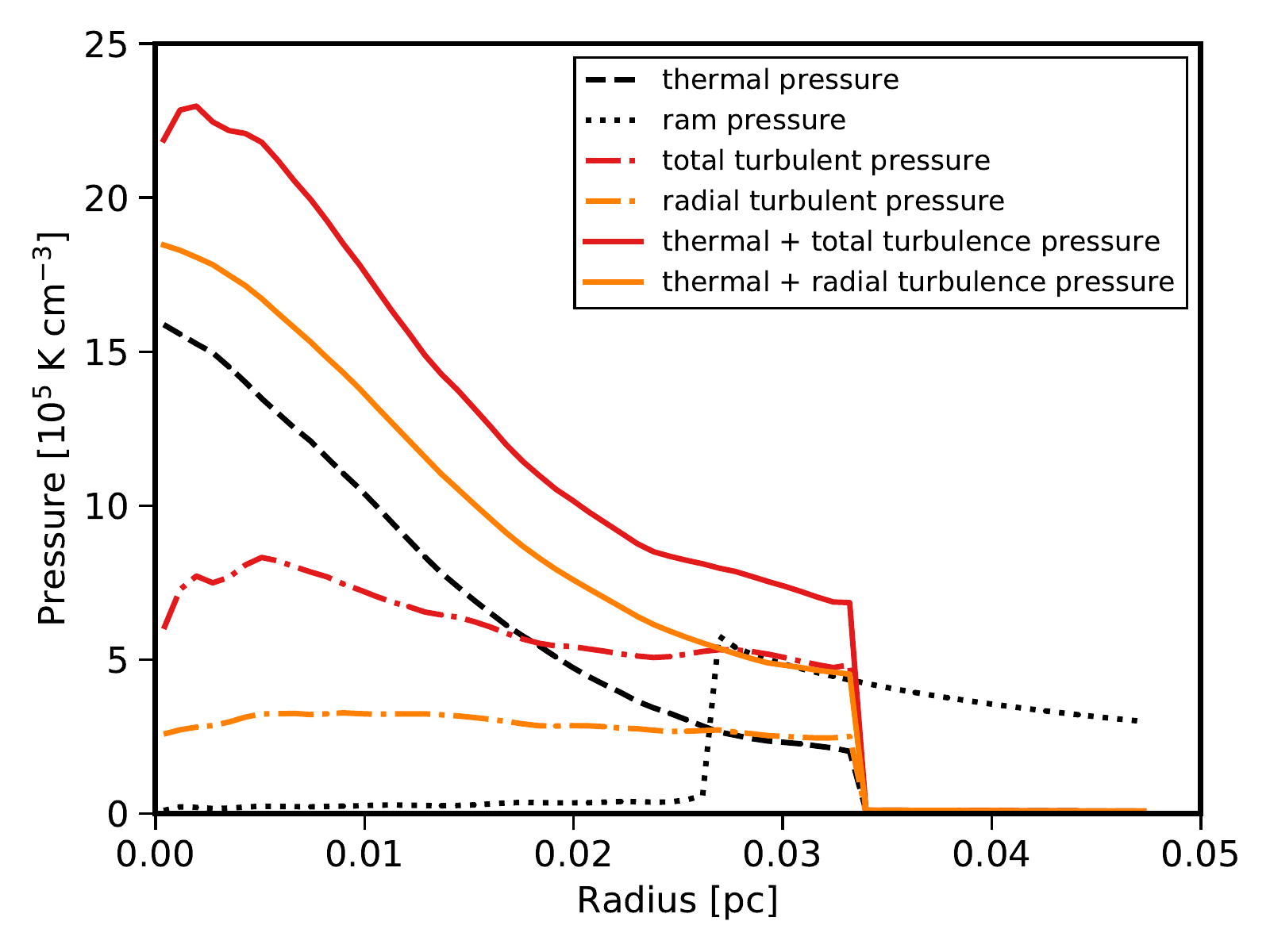}
     \caption{The different pressure contributions averaged in radial bins
     of our fiducial filament with an accretion flow of Mach 6.0 and an
     accretion rate of 16.8 M$_\odot$ pc$^{-1}$ Myr$^{-1}$ measured at 0.5
     Myrs. The black dashed line shows the thermal pressure of the filament
     gas and the black dotted line the ram pressure in the accretion flow.
     The turbulent pressure is given by the dashed-dotted lines and the sum
     together with the thermal pressure to show the total filament pressure
     by the solid lines: for the total velocity dispersion in red and for
     the radial velocity dispersion in orange.}
     \label{fig:rad_p}
   \end{figure}

   In order to determine why there is no pressure support, we analyse
   the pressure profile of our high resolution fiducial filament as we
   did for the turbulent Mach number in \autoref{fig:rad_vkin}. We
   distinguish between cells that are part of the filament and others
   which trace the accretion flow by using the same method of determining
   the filament radius by the jump in radial velocity as mentioned above.
   As the filament radius is not uniform as can be seen in the
   example cuts in \autoref{fig:rad_vel}, there is an overlap region where
   cells of both regions are present. We calculate the respective pressures
   with cells of the respective region and cells only contribute
   to the respective pressure on an individual basis. For the turbulent
   pressure we use only cells tracing
   the filament and determine the average density and kinetic energy in
   each bin $i$ from which we calculate the pressure as
   $\left<\rho_i\right>\sigma_i^2$ and then average over all slices. For
   the ram pressure we determine the average density and radial velocity
   in the accretion flow and calculate the ram pressure as
   $\left<\rho_i\right>\left<v_i^\text{rad}\right>^2$ and then
   average over all slices. The resulting pressure components are shown
   in \autoref{fig:rad_p}. The thermal pressure is given by the black dashed
   line, the ram pressure by the black dotted line, the total and radial
   turbulent pressure by the red and orange dashed-dotted line respectively
   and the combined thermal plus turbulent pressure by the respective solid
   line. The overlap region is illustrated by the dotted ram pressure
   line extending into the filament pressure over a broad range around the
   mean radius of about 0.03 pc.

   We already showed in the last subsection that the radial turbulent
   motions provide the hydrostatic equilibrium together with the ram
   pressure. This can be seen also in the pressure directly, where the
   combined thermal plus radial turbulent pressure given by the orange solid
   line exactly balance the ram pressure in the overlap region. In contrast,
   the total turbulent pressure provides a pressure which is too large for
   an equilibrium. Moreover, one can see that the accretion driven turbulence,
   given by the dashed-dotted lines, distributes itself over the
   filament in a way that the turbulent pressure component is constant
   throughout the filament. In the context of large-scale in
   comparison to small-scale turbulence discussed in the introduction, it
   seems to resemble the classical Kolmogorov model as it does not form
   overdense regions of different levels of turbulence. This leads to the
   fact that the pressure profile of the filament is only shifted to an
   overall larger pressure by a constant value due to turbulence
   as can be seen in the figure by comparing the dashed black and
   the solid orange line. One can interpret this as a shift in the
   isothermal equation of state to include a constant offset:
   \begin{equation}
     p = \rho c_s^2 + p_0,
   \end{equation}
   which does not change the solution of the isothermal, cylindrical
   Lane-Emden equation, as the hydrostatic equilibrium depends only on the
   gradient of the pressure. The scaling of the profile does not change and
   thus the maximum line-mass remains the same.

   \begin{figure}
     \includegraphics[width=1.0\columnwidth]{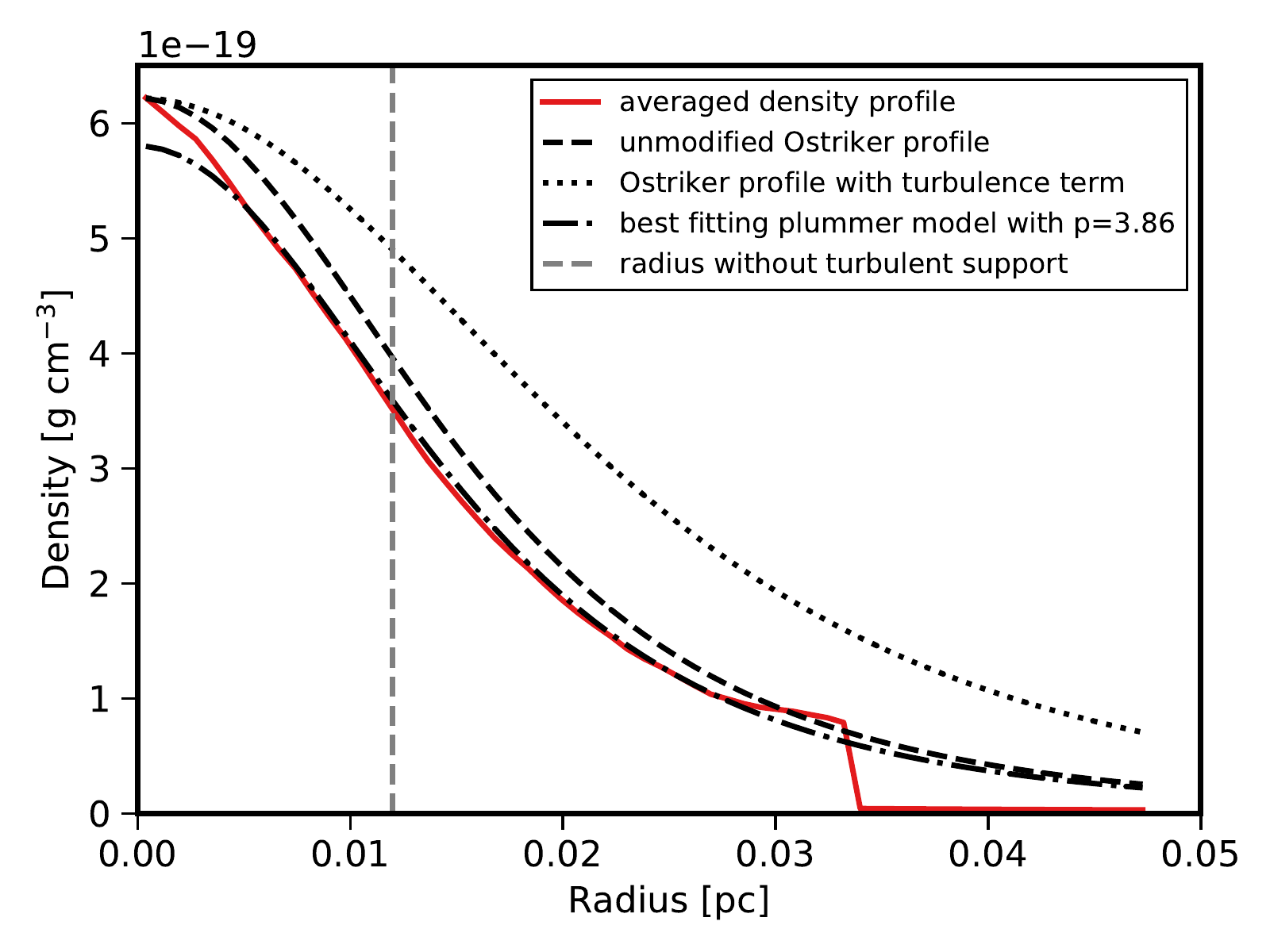}
     \caption{The measured density profile constructed from averaged radial
     bins of our fiducial filament at 0.5 Myrs, the timestep when the radius
     reaches its maximum extent. The black dashed line shows the measured
     density of the filament gas while the dashed and dotted gray lines show
     the corresponding density profiles using the same central density as
     the measured profile. The dashed line is given by the unchanged
     Ostriker scaling and the dotted line is given by including an additional
     pressure contribution. The dashed-dotted gray line shows the best
     fitting plummer model with an exponent of -3.86. We also show the radial
     extent we would expect without any turbulent pressure contribution as
     given by the dashed black line (see also \autoref{fig:f_rad}).}
     \label{fig:rad_rho}
   \end{figure}

   This effect can also
   be seen in the radial density profile in \autoref{fig:rad_rho}. Here we
   show the measured density averaged over all slices along the filament as
   the solid red line. Additionally, we overplot the different expected
   density profiles, once with a turbulent pressure contribution as given in
   \autoref{eq:lmturb} shown as the dotted line and once without an
   additional pressure as given by the dashed line. For both cases we
   use the central density to normalise the scale height. One can see that
   the measured profile follows the unmodified density almost perfectly
   albeit there is a slight offset due to what seems to be a small
   over-density in the centre of the filament skewing the resulting profile.
   In order to check if the accretion changes the slope of the profile, we
   also fit a plummer profile to the measured density which is free to
   vary in the central density, in the scale height and in power. The best
   fitting model is given by the dashed-dotted line and shows an
   exponent of 3.86 which is very close to the analytic value of 4. Thus,
   the profile is softened barely by the accretion process. As the
   boundary pressure of the filament is larger, it extends further into
   the surrounding medium. This is demonstrated by the vertical
   dashed black line which shows the extent of the filament we would expect
   lacking any internal turbulent pressure. Note, that turbulence thus can
   influence the absolute value of the scale height by setting the central
   density via the radial extent, but it does not change the
   general scaling of the profile and therefore is not added as isotropic
   pressure contribution to the sound speed.

   \section{Theoretical implications for core formation}
   \label{sec:core}

\begin{figure}
     \includegraphics[width=1.0\columnwidth]{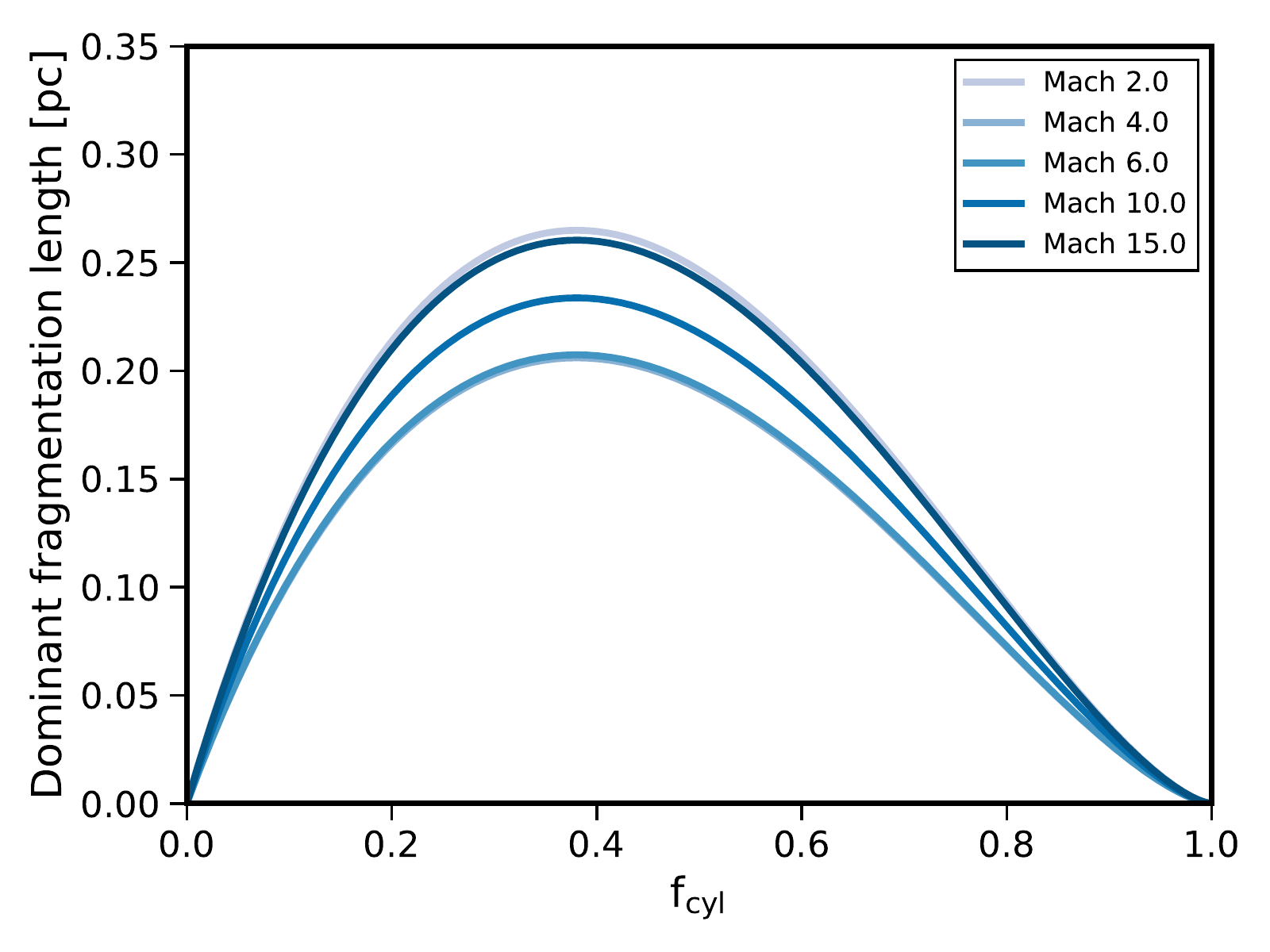}
     \caption{The dominant fragmentation length as a function of the line-mass
     for the same mass accretion rate of 16.8 M$_\odot$ pc$^{-1}$ Myr$^{-1}$
     but for different inflow Mach numbers.}
     \label{fig:l_dom}
   \end{figure}

   In all our simulations independent of the inflow Mach number,
   the radial velocity dispersion at the filament boundary amounts to about
   0.85 times the total equilibrium velocity dispersion of the non
   self-gravitational case for which a functional form can be found
   in the appendix. Thus, we can calculate the theoretical radius
   and central density of the filament at every line-mass and
   therefore we can make predictions on the fragmentation length and
   time-scales of cores forming in an accreting filament using the
   gravitational fragmentation model. This model was successfully
   applied to explain several observed core distances \citep{jackson2010,
   miettinen2012, busquet2013, beuther2015, contreras2016, heigl2016,
   kainulainen2016} however it is not able to explain all observations
   \citep{andre2010, kainulainen2013, takahashi2013, lu2014, wang2014,
   henshaw2016, teixeira2016, kainulainen2017, lu2018, palau2018,
   williams2018, zhou2019}. It predicts that small density perturbations
   in the linear regime along the filament axis of the form:
   \begin{equation}
     \rho(r, x, t) = \rho_0(r) \left(1 + \epsilon \exp(ikx -i\omega t)\right)
   \end{equation}
   will grow for values of $k$ where the dispersion relation $\omega^2(k)$
   is negative. Here $\rho_0$ is the unperturbed initial density,
   $k = 2\pi/\lambda$ is the wave vector with $\lambda$ being the
   perturbation length, $x$ is the filament axis, $\omega = 1/\tau$ is the
   growth rate with $\tau$ being the growth timescale, $t$ the time variable
   and $\epsilon$ the perturbation strength. The fastest growing, or dominant,
   fragmentation length scale $\lambda_\text{dom}$ as well as the growth
   timescale of the dominant mode $\tau_\text{dom}$ depend on the current
   line-mass as well as the current central density of
   the filament and are given by the pre-calculated \citep{nagasawa1987}
   and interpolated values in \citet{fischera2012}, shown by their table
   E.1. We use these values to determine the length scale of the fastest
   growing mode at every line-mass for the same mass accretion
   rate but for different inflow Mach numbers as shown in \autoref{fig:l_dom}.
   As one can see, the dominant fragmentation length changes over the
   evolution of the line-mass. At the boundary values it vanishes to zero
   and it has a maximum at about $\fcyl=0.4$. The figure is self-similar
   for different mass accretion rates, with a lower rate leading to a
   larger dominant fragmentation length. For a constant accretion rate,
   the fragmentation length does not
   vary much for different inflow Mach numbers. Only for large and for very
   low inflow Mach numbers, the fragmentation length is slightly larger. As
   the dominant fragmentation length is constantly changes as $\fcyl$
   grows, it is hard to make predictions of what will be the final
   distance between forming cores. But the curves have a maximum which
   allows us to make a
   prediction about the minimum number of cores that will form. For instance,
   a filament with an inflow Mach number of 4.0 and a length of 0.2 pc will
   form at least one core. As soon as the first core forms, the further
   evolution of the filament is also influenced by the gravitational
   attraction of the core. This makes the formation of additional cores
   even more unpredictable.

   We can get some further constraints form calculating the core
   growth time. As the filament grows in mass, we assume that cores are
   seeded at each line-mass on the temporary dominant wavelength with a
   perturbation strength of 0.09 given by the standard deviation of observed
   line-masses in the study by \citet{roy2015}. In order for cores to be
   observed at a certain distance, the cores have to grow faster than the
   filament itself lest their local line-mass enhancement is overtaken by
   the overall line-mass growth of the filament. The limit of this growth
   is the critical line-mass where the cores locally collapse radially and
   which value they have to reach before the overall filament in order to
   be observable as cores. Therefore, we compare the timescale of the
   overall filament to reach the critical line-mass via accretion to the
   timescale the cores would need to reach this value if they would continue
   growing on the dominant timescale. This means we solve the equation of
   the growth of the line-mass for the time $t$ where the line-mass
   enhancement reaches a value of one:
   \begin{equation}
     \fcyl^\text{max}(t) = \fcyl^0\left(1 + \epsilon \exp(t/\tau_\text{dom})\right) = 1.0,
   \end{equation}
   where $\fcyl^0$ is the unperturbed line-mass at the beginning of the
   growth of the respective fastest growing mode. One also has to assume
   a perturbation strength $\epsilon$ which we set to 0.09 as mentioned
   above. Note, that the dominant timescale grows to longer values as the
   filament evolves as it is only dominant for the initial line-mass where
   the cores start growing. Therefore, our calculation represents the most
   optimistic case and the time to reach the critical line-mass is only a
   lower boundary.

\begin{figure}
     \includegraphics[width=1.0\columnwidth]{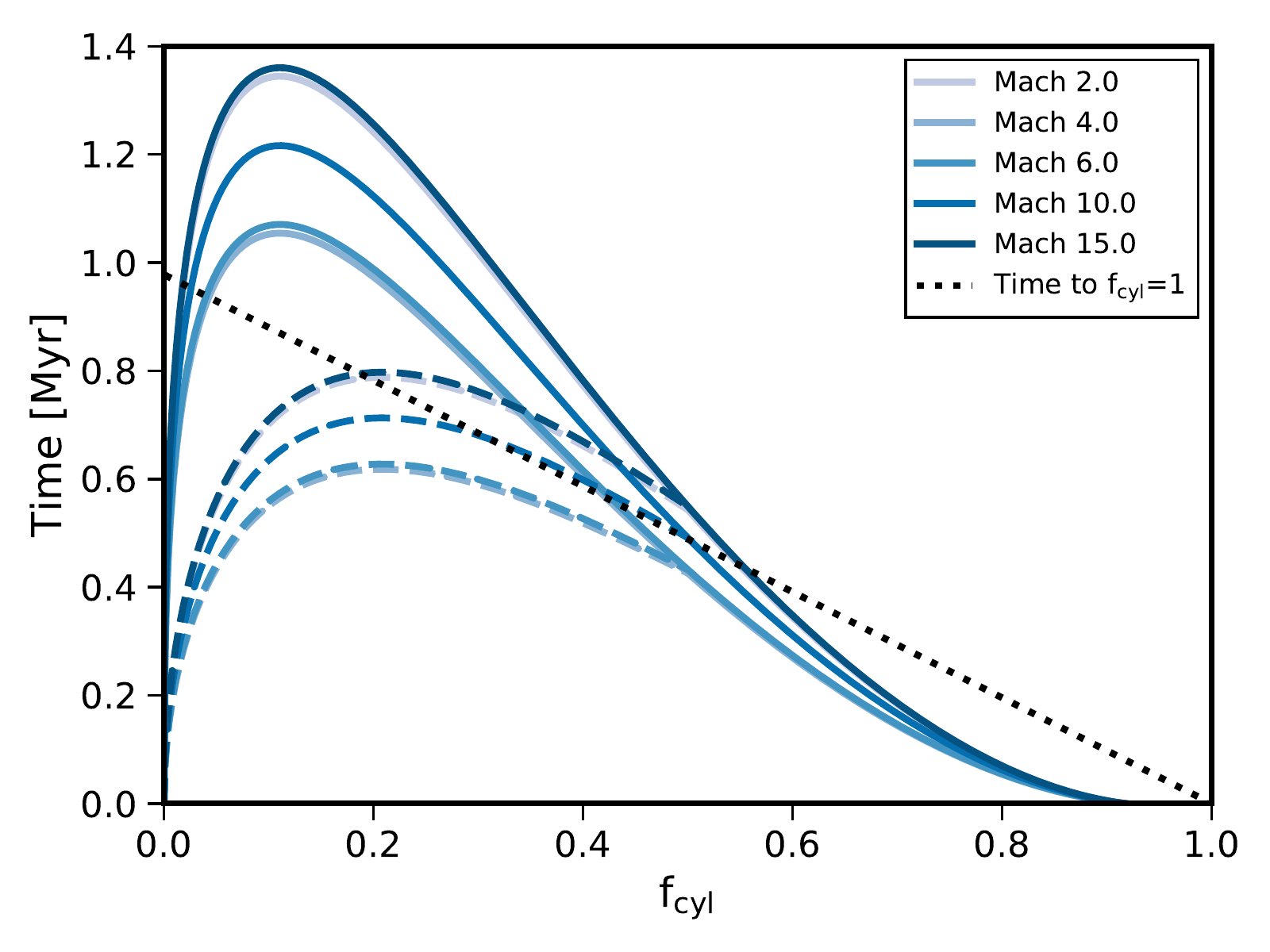}
      \caption{The dominant fragmentation time compared to the collapse time
      of a filament as a function of the line-mass. The dashed line shows
      the remaining time to reach a value of $\fcyl=1.0$ compared to the time
      it takes a growing core to reach the same value at each line-mass.
      Only at values of $\fcyl\approx0.7$ and above, cores have time to form
      before the filament collapses.}
      \label{fig:t_dom}
   \end{figure}

   The result of this calculation is shown in
   \autoref{fig:t_dom}. The filament is accreting mass at a constant rate,
   thus the time for to reach the maximum line-mass is decreasing linearly
   as shown by the black dotted line. As one can see, the growth timescale
   of a core is larger than the filament collapse time for the majority of
   its evolution. The dominant growth timescale is shorter for large central
   densities as is the case at very low and high values of $\fcyl$ where the
   filament is centrally concentrated. The upper value of $\fcyl$ where the
   growth time curves intersect the collapse timescale of the filament is
   approximately where we also typically observe core formation in our
   simulations. For lower values of $\fcyl$ we never see any core
   formation occurring. We do see local overdensities on very small
   length scales similar to random noise but no real core forms. As the
   dominant length scale changes over time, any pre-existing overdensity
   is washed out. This changes for high line-masses as here the cores form
   local overdensities which large gravitational attraction suppresses the
   further change of the dominant mode. The timescale to reach the
   critical line-mass is very small and the collapse of the core is
   irreversible.

   However, filaments of seed line-masses of values lower than 0.5
   theoretically could never reach the critical line-mass in the linear
   regime as the linear model allows at most an increase of the local
   line-mass of a factor of two. There are non-linear effects where it
   could accrete mass from all over the filament \citep{heigl2016} but
   here we assume it continues to grow in the linear regime and the
   additional mass is provided by the filament accretion. Nevertheless, we
   also show the timescales to reach the theoretical maximum local line-mass
   given by the dashed lines. Note, that if we calculate the timescale the
   overall filament needs to reach the same value from the same initial
   line-mass, it is always faster than the cores themselves.

   As in the case of the fragmentation length, the growth timescale
   is self-similar. A larger mass accretion rate only shortens the growth
   time as well as the time for the filament to reach the maximum line-mass
   in the same manner. Note, that this relation therefore also holds true
   for an increasing inflow velocity due to a growing line-mass. This
   implies that if cores are observed in an accreting filament, it
   is more likely to have a line-mass closer to the maximum line-mass. From
   the results of \citet{heigl2018_2} which show that cores forming in high
   line-mass filaments lead to a reduced filament radius at the position of
   the core, we also expect the cores to have a thinner morphology than the
   filament itself.

\section{Discussion and conclusions}
\label{sec:discussion}

   This work presents a numerical study on accretion driven turbulence in
   filaments. Together with the results of paper I, we have shown
   that, depending on inflow velocity, accretion flows with realistic inflow
   velocities and observed mass accretion rates can drive turbulent motions
   ranging from subsonic to supersonic velocities. The major difference to
   the former study without gravity is a limited filament radius and an
   associated decaying velocity dispersion. Moreover, accretion driven
   turbulence leads to a radial profile of the velocity dispersion which
   is anti-correlated to the density profile, thus resulting in a constant
   turbulent pressure throughout the filament and therefore to a lack of
   turbulent pressure support. However, our model relies
   on several assumptions.

   First of all, our simulations lack magnetic fields which could suppress
   turbulent motions. Although magnetic fields are thought of channelling
   accretion flows along striations, density enhancements perpendicular to
   the filament \citep{goldsmith2008, palmeirim2013, cox2016}, they have
   been shown to stabilise filament against fragmentation depending on the
   field configuration
   \citep{stodolkiewicz1963, nagasawa1987, 2gehman1996, 2fiege2000}. They
   can act as an additional pressure support and also suppress
   motions perpendicular to the field lines. The effects of magnetic fields
   therefore will be explored in a future paper.

   Furthermore, while we do include an initial density perturbation in order
   to break the symmetry, our accretion flow is very smooth. It could be that
   accretion is better treated by the infall of clumpy material or even
   with initial turbulent velocity distributions as in \citet{clarke2017}.
   Filaments do not form in isolation and driven turbulent box simulations
   show filaments forming as transient entities \citep{federrath2016}.
   To that end, large scale simulations with realistic inflows in an molecular
   cloud environment are needed which are out of the scope of this work.

   Observations of the massive filament DR21 show an increasing velocity
   dispersion toward the central axis of the filament \citep{schneider2010}.
   Our models shows a decreasing velocity dispersion towards the
   centre of the filament. However, one has to take projection effects
   into account. While mock observations of our models do not show
   an obvious increase, we also do not see any systematic drop of velocity
   dispersion towards the centre of the filament either. However, it could
   be that the potential of a comparison is limited as DR21 is supercritical
   and probably in a state of radial collapse.

   Nevertheless, all our simulations show the lack of turbulent
   pressure support against radial collapse independent of inflow Mach
   number. This constitutes an interesting case where turbulence does
   not act as an additional pressure. We can summarise our findings as
   the following:

   \begin{itemize}
   \item A smooth radial accretion onto filaments drives turbulent motions
     which are radially dominated and decay over time.
   \item The turbulent pressure has a radial profile which is anti-correlated
     to the density as the low density outer layers are easier to stir.
   \item This leads to a constant turbulent pressure component which
     does not add radial stability as the stability relies on pressure
     gradients.
   \item We predict that cores usually form for higher line-mass in
     accreting filaments ($\fcyl \geq 0.5$) as only then their growth is
     fast enough to outpace the collapse of the entire filament.
   \end{itemize}

\section*{Acknowledgements}

   We thank Alvaro Hacar and the whole CAST group for helpful
   comments and discussions. Furthermore, we thank the referee for
   improving the quality of the paper. AB, MG and SH have been
   supported by the priority programme 1573 "Physics of the Interstellar
   Medium" of the German Science Foundation and the Cluster of Excellence
   "Origin and Structure of the Universe". The simulations were run using
   resources of the Leibniz Rechenzentrum
   (LRZ, Munich; linux cluster CoolMUC2).



\bibliographystyle{mnras}
\bibliography{Turb.bib}



\appendix

\section{Non self-gravitational radius evolution revisited}
\label{sec:equilibrium}

   While we found a linear relation between the inflow velocity
   and the equilibrium velocity dispersion in paper I, we could not
   find a reasonable explanation for the offset of the linear fit.
   As we have refined our measurement techniques and updated our
   model, we want to discuss the implications of this study on our
   previous work.

   \subsection{Radial evolution without self-gravity}

   As the density inside the filament is constant in the non
   self-gravitational case, we formerly modeled the radius by
   the mass accretion rate (\autoref{eq:mdot}) assuming that the mean density
   inside the filament is given by the exterior density times the square of
   the Mach number in an isothermal shock:
   \begin{equation}
     \left<\rho\right>_\text{fil} = \dext \mathcal{M}_a^2.
   \end{equation}
   The mass accretion rate together with the total mass of the filament
   \begin{equation}
     M = \pi R^2 L \left<\rho\right>_\text{fil}
       = \pi R^2 L \dext \mathcal{M}_a^2
       = \pi R L \rho_0 R_0 \mathcal{M}_a^2,
   \end{equation}
   then leads to a radius evolution given by
   \begin{equation}
     R(t) = \frac{2 c_s^2 t}{v_a}.
     \label{eq:radwog}
   \end{equation}
   While we found a good agreement, this does not take into account the
   effect of turbulence on the radius. A better model is to consider the
   pressure equilibrium at the filament boundary, analogous to
   \autoref{eq:pwg}, supported internally by the turbulent pressure
   $\left<\rho\right>_\text{fil}\sigma^2$ and by the ram pressure
   $\dext v_a^2$ on the outside:
   \begin{equation}
     \left<\rho\right>_\text{fil} \left(c_s^2 + \sigma^2 \right)
     = \dext \left(c_s^2 + v_a^2 \right).
   \end{equation}
   This gives the internal mean density
   \begin{equation}
     \left<\rho\right>_\text{fil} = \dext\frac{\left(1 + \mathcal{M}_a^2\right)}{\left(1+\mathcal{M}_t^2\right)}
   \end{equation}
   which leads to the radius evolution of
   \begin{equation}
     R(t) = 2v_at\Mratio.
     \label{eq:radwogturb}
   \end{equation}

   \begin{figure}
     \includegraphics[width=1.0\columnwidth]{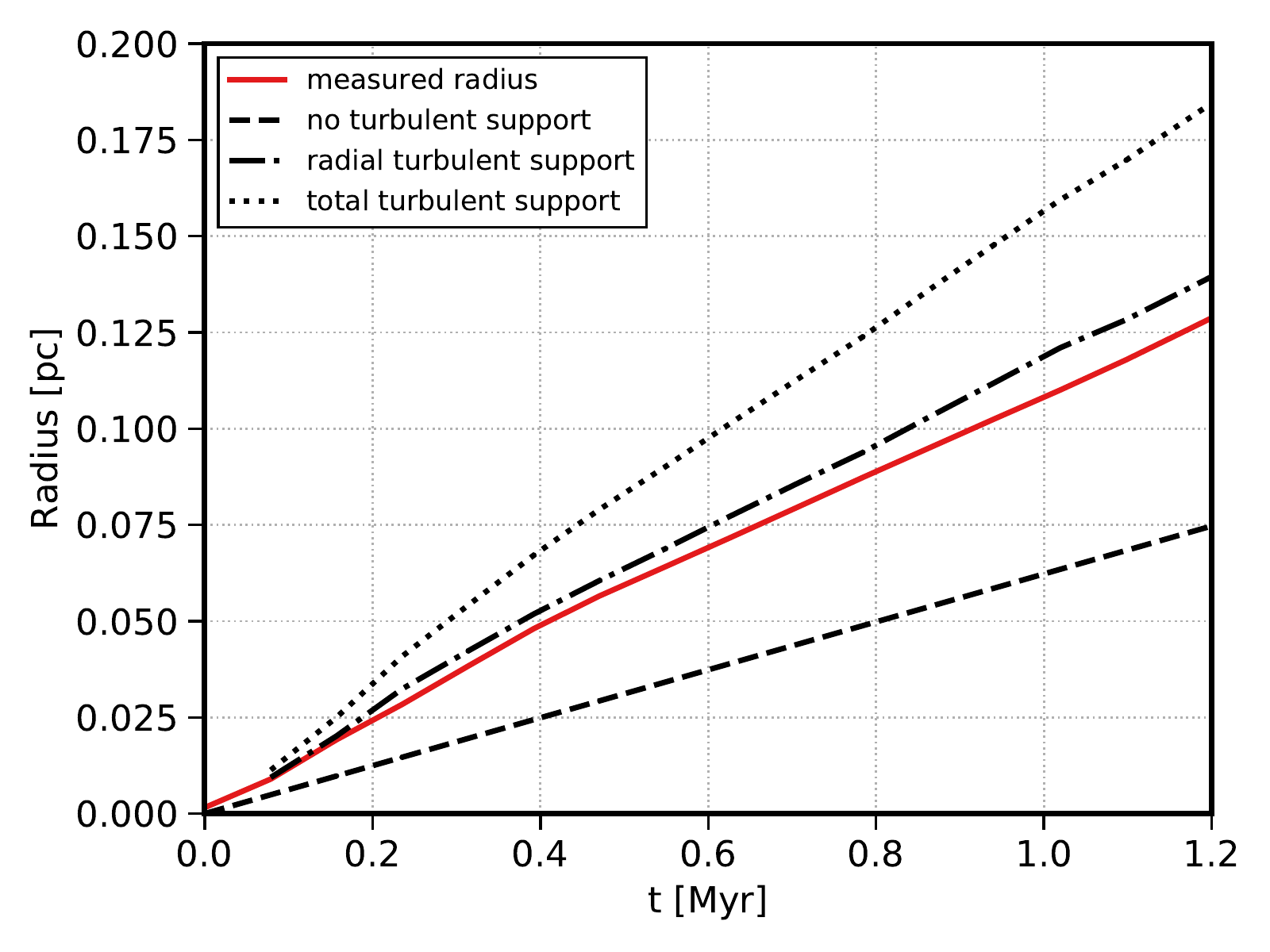}
     \caption{Radial evolution of a filament with a Mach 6 inflow without
     gravity. The solid black line is the measured radius at each time step.
     The theoretical predictions are given by \autoref{eq:radwogturb} with
     the full, radial and no turbulent support is given by the dotted,
     dashed-dotted and dashed line respectively.}
     \label{fig:t_rad}
   \end{figure}

   \begin{figure*}
     \includegraphics[width=2.0\columnwidth]{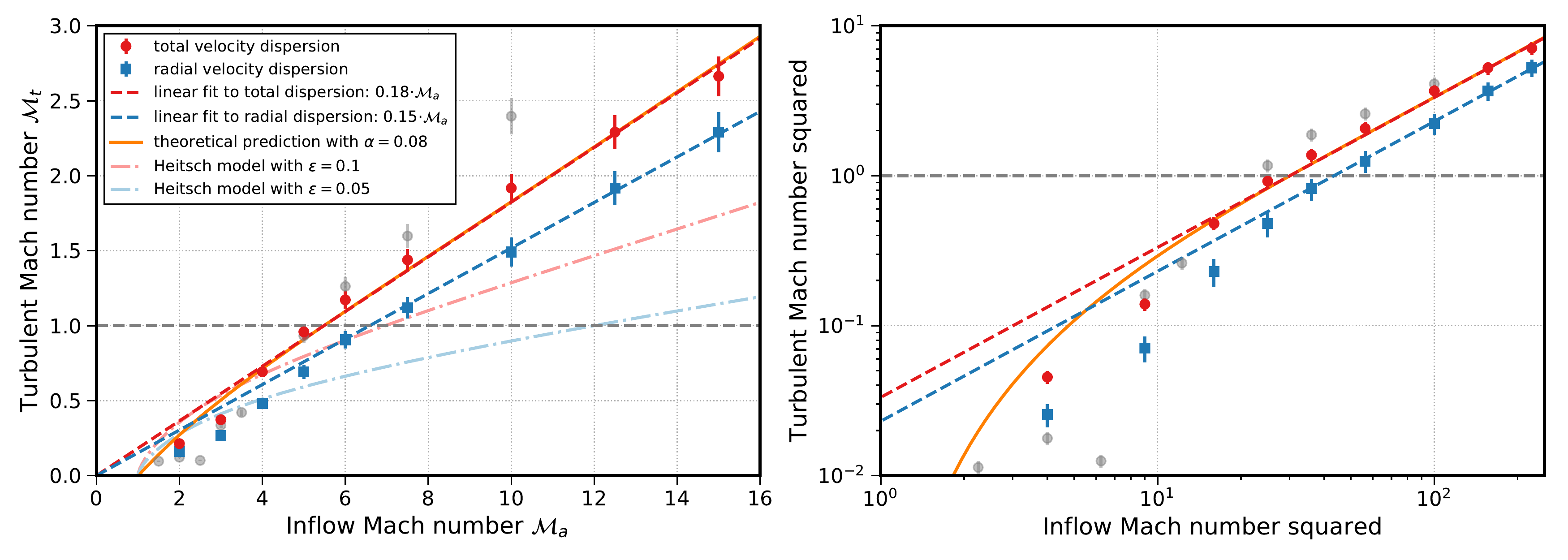}
     \caption{Equilibrium values of accretion driven turbulence in dependence
     of the inflow Mach number. Both plots show the same data points. On the
     left hand side we directly plot the values which we calculate from the
     kinetic energy. The gray data points are the values of the density weighted
     standard deviation we used in paper I. On the right hand side we show
     the same data points with squared values in log-log scale.}
     \label{fig:Ma_Mt}
   \end{figure*}

   \noindent As we have shown in \autoref{fig:t_ekin_split} the turbulence inside the
   filament is not isotropic. Thus, it is not necessarily the total
   velocity dispersion which supports the internal pressure. Furthermore,
   it is important to note that the hydrostatic equilibrium is given by
   the turbulent pressure at the boundary. As the non self-gravitating
   velocity dispersion has a similar radial profile as shown for the
   self-gravitating case in \autoref{fig:rad_vkin},
   we would need to use the boundary velocity dispersion value in
   \autoref{eq:radwogturb}. However, for the non self-gravitational case,
   the boundary velocity dispersion is very similar to the velocity
   dispersion calculated from the total filament as it does not have a
   density profile. Therefore, we do not see an anti-correlation between
   density and velocity dispersion and the latter is then dominated by the
   largest value which is at the boundary. We plot the radial evolution as
   solid red line in \autoref{fig:t_rad} together with the expected evolution
   including no turbulent support as dashed, only radial support as
   dashed-dotted and total turbulence support as dotted black line. Compared
   to paper I, we improved our measurement method of
   the radius by not using the largest density jump but by using the same
   methodology as we do for the velocity dispersion which distinguishes
   filament material from the surrounding by using the drop in radial
   velocity. We found that the density inside the filament can vary
   substantially and the largest density jump does underestimate
   the filament radius. Moreover, sometimes the filament can be deformed,
   having an more elliptical formed cross-section. We reduce the impact
   of this effect by using the mean radius of two perpendicular cuts through
   the center of the filament. The expected evolution follows from
   \autoref{eq:radwogturb} together with the measured velocity dispersion
   at every time-step. Note that the no turbulent support case is
   indistinguishable from the isothermal shock radial evolution of
   \autoref{eq:radwog} for large inflow Mach numbers.

   As one can see, the radius is best fitted by a support of the radial
   component of the velocity dispersion. This shows that only radial
   motions of the filament are important in setting the boundary pressure
   against the radial inflow.

   \subsection{Equilibrium velocity dispersion}

   We now want to use the information about the non self-gravitational
   radius evolution to explain the linear relationship we found in
   paper I between the inflow velocity and the created turbulence.
   We measure the equilibrium value which we plot as red data points
   in \autoref{fig:Ma_Mt}. The left panel shows the direct relationship
   between the inflow velocity and the generated turbulence. On the right
   hand side we plot the same values but squared in order to compare
   the inflow energy to the turbulent energy. Additionally, we show the
   values of the radial turbulence as blue squares. The error bars
   indicate that not all simulations give the same equilibrium
   number but different seeds in the random density distribution result in
   slightly different equilibrium values with a spread of about 10\%.
   However, this error does not enter our fit to the datapoints.
   In gray, we also plot the data values of paper I which where
   calculated using the density weighted standard deviation. One can see
   that for low values the differences are not huge but they become
   increasingly larger for higher inflow Mach numbers, as discussed at the
   beginning of \autoref{sec:simulations}. Compared to paper I, we also
   extend the range of inflow velocities to Mach 15 which is already much
   greater than the expected values but serves as an good upper limit.

   The reason why in general there is an equilibrium level in the velocity
   dispersion in the non-gravitational case is due to
   \autoref{eq:equilibrium}. If the radius is growing linearly with time as
   shown in \autoref{eq:radwogturb} and if the equilibrium has been
   established, the dissipation rate is constant in time (\autoref{eq:ediss}):
   \begin{equation}
     \dot{E}_d \approx \half \frac{M(t) \sigma^3}{L_d(t)} =
               \half \frac{\dot{M} t \sigma^3 \left(1+\mathcal{M}_a^2\right)}{4 v_a t\left(1+\mathcal{M}_t^2\right)} =
               \frac{\dot{M}\sigma^3 \left(1+\mathcal{M}_a^2\right)}{8 v_a \left(1+\mathcal{M}_t^2\right)}.
   \end{equation}
   If the velocity dispersion is greater than the equilibrium value, the
   excess is dissipated away, if it is lower, less energy is dissipated.
   Thus it will settle at a value where the dissipation is constant.

   In order to compare to theoretical models we first discuss the
   prediction by \citet{heitsch2013}. It assumes that one can express
   the dissipation rate as a constant fraction of the constant energy inflow
   (\autoref{eq:epsilon}):
   \begin{equation}
     \epsilon = \left| \frac{\dot{E}_d}{\dot{E}_a} \right|.
   \end{equation}
   Note that $\epsilon$ can only be constant in time for a constant mass
   accretion rate if the dissipation rate is constant in time. For a
   non-linear radial evolution this is not the case. Furthermore, the model
   assumes that $\epsilon$ is independent of the inflow velocity which is
   not necessarily true as the fraction of accreted energy converted to
   turbulent motions can change with the inflow velocity. Nevertheless, in
   the non-gravitational case \autoref{eq:heitsch} transforms to
   \begin{equation}
     \frac{\mathcal{M}_t^3}{\left(1+\mathcal{M}_t^2\right)} = \frac{4\epsilon\mathcal{M}_a^3}{\left(1+\mathcal{M}_a^2\right)}.
     \label{eq:heitschwog}
   \end{equation}
   As turbulence is generated in oblique shocks on the surface of the filament,
   we need supersonic inflow motions. Below an inflow velocity of Mach 1.0 we do
   not generate turbulent motions or even form a pressure bound filament.
   Therefore, we shift the zero point of \autoref{eq:heitschwog} to an inflow
   velocity of Mach 1.0 by effectively applying the transformation
   $\mathcal{M}_a' = \mathcal{M}_a-1$. Note that this transformation only
   affects the energy accretion term and not to the evolution of the radius.
   Thus, the equation is now:
   \begin{equation}
     \frac{\mathcal{M}_t^3}{\left(1+\mathcal{M}_t^2\right)} = \frac{4\epsilon\mathcal{M}_a'^2\mathcal{M}_a}{\left(1+\mathcal{M}_a^2\right)}.
   \end{equation}
   A realistic estimate of $\epsilon$ is expected to lie between 5\% and 10\%
   \citep{klessen2010} which we plot as the dashed-dotted light blue and red
   lines respectively. As one can see the curves do not fit the measured points
   and fit even worse if we do not apply the transformation. The shape of the
   curve cannot be matched to the data points even if we fit different values
   of $\epsilon$. This leaves us with the conclusion that $\epsilon$ is only
   a constant in time for a certain inflow velocity but varies with the inflow
   velocity.

   As we clearly cannot apply the model by \citet{heitsch2013}, we try to
   fit \autoref{eq:equilibrium} directly. If we insert the evolution of the
   non self-gravitational radius, it transforms to
   \begin{equation}
     \mathcal{M}_t^2 = \alpha \mathcal{M}_a'^2 - \frac{\mathcal{M}_t^3 \left(1+\mathcal{M}_a^2\right)}{4 \mathcal{M}_a \left(1+\mathcal{M}_t^2\right)}.
   \end{equation}
   We fit this relation in \autoref{fig:Ma_Mt} as solid, orange line and
   get the best fitting value of $\alpha = 0.085$. As one can see it follows
   the data points well and has the same scaling for large $\mathcal{M}_a$.
   Only for low inflow Mach numbers there is some discrepancy where the data
   points lie not exactly on the relation. Therefore, our model seems to
   reasonably explain the connection of accretion driven turbulence an
   inflow Mach number.

   Nevertheless, our simulations show that most of the inflow energy
   is lost. In the accretion process only $8.5\%$ of the energy is kept in
   the shock phase and if we convert the linear fit to the total velocity
   dispersion to an energy relation, it indicates that from this value only
   only about $0.18^2 = 3.2\%$ of the energy is retained in turbulent motions
   at equilibrium for large inflow Mach numbers. The energy difference
   between these two values is lost in the continuous dissipation inside the
   filament.



\bsp	
\label{lastpage}
\end{document}